\documentstyle[preprint,aps,eqsecnum]{revtex}
\begin{document}

\title{Exciton-exciton interactions in quantum wells.
Optical properties and energy and spin relaxation.}

\author{S. Ben-Tabou de-Leon and B. Laikhtman}

\address{
Racah Institute of Physics, The Hebrew University, Jerusalem 91904,
Israel}

\maketitle

\begin{abstract}
The gas of interacting excitons in quantum wells is studied. We obtain the
Hamiltonian of this gas by the projection of the electron-hole plasma
Hamiltonian to exciton states and an expansion in a small density. Matrix
elements of the exciton Hamiltonian are rather sensitive to the geometry of
the heterostructure. The mean field approximation of the exciton Hamiltonian
gives the blue shift and spin splitting of the exciton luminescence lines.
We also write down the Boltzmann equation for excitons and estimate the
energy and spin relaxation time resulting from the exciton-exciton
scattering. Making use of these calculations we succeeded to explain some
recent experimental results which have not been explained so far.
\end{abstract}


\section{Introduction}
\label{int}

Optical properties of dense exciton gas in quantum wells had been studied
intensively both experimentally and theoretically in the past decade.
Originally, attempts to obtain dense exciton gas were motivated by a desire
to reach Bose - Einstein condensation of excitons. However, the fascinating
observed physical phenomena and the promising potential for device
application had made the exciton gas a very interesting system by itself.

Recent experiments have stressed the importance of exciton-exciton
interaction of two-dimensional (2D) excitons in quantum wells.
Exciton-exciton interaction affects exciton photoluminescence, breaks the
symmetry between excitons with different spin components, contributes to
depolarization of the exciton gas, and controls spin and momentum relaxation
in the exciton system. A thorough investigation of exciton-exciton
interaction is crucial for understanding of a number of experimental results
in exciton gas.

Blue shift of the exciton luminescence line with growing exciton density was
reported in single quantum wells \cite{Hulin1,Hulin2}, multiple quantum
wells\cite{Dareys93,Amand94} and coupled quantum wells where electrons and
holes are spatially separated \cite{Butov,Snoke}. The blue shift is
attributed to the net exciton-exciton repulsion interaction in quantum wells
\cite{Schmitt-Rink,Tejedor96,Ciuti98}.

Other interesting interaction induced phenomena are related to exciton spin.
Here we use the term spin for the projection of the total angular momentum
of the exciton to the direction perpendicular to quantum well plane (the
growth direction). In quantum wells size quantization leads to an energy
separation between heavy and light holes. Therefore one can treat the ground
state exciton as a bound state of a conduction electron with spin $\pm1/2$
and a heavy hole with spin $\pm3/2$. So, in the ground state, i.e., s-state
exciton, the spin can take the values $\pm 1$ and $\pm 2$. Total momentum
conservation allows only states with spin projection $\pm 1$ to be optically
active in single photon experiments.

Spin +1 and spin -1 excitons can be created independently by pumping
circular polarized light. A spontaneous energy splitting between the two
different components of spin-polarized exciton gas in zero magnetic field was
observed in multiple quantum wells \cite{Damen91,Vina96,Jeune98,Vina99} and
in coupled quantum wells \cite{Mendez}. The splitting increased with the
exciton density and decreased with the separation of the electron and hole
induced by external electric field in coupled quantum wells. The splitting
was explained by the exciton-exciton exchange interactions \cite{Tejedor96}.

At high exciton densities the decay of the exciton luminescence cannot be
characterized by one decay time. Initially the decay is very fast and then
it is followed by a slower relaxation.\cite{Shah,Baylac95,Amand97} Similar
results were obtained for the decay of luminescence
polarization.\cite{Baylac95,Amand97} A very careful analysis of the
luminescence intensity and polarization by Baylac, Amand {\it et
al}.\cite{Baylac95,Amand97} and two-pulse experiments performed by Le Jeune
{\it et al}.\cite{Jeune98} proved that the short decay time can be explained
only by exciton-exciton scattering. In dense exciton gas exciton-exciton
scattering can also be the leading mechanism of momentum relaxation.
\cite{Wang}

The active investigation of exciton-exciton interaction and many new
experimental results makes it difficult to overestimate the importance of
theoretical description of this interaction. Such a description, however,
encounters significant difficulties even when the exciton density is small
and the interaction between excitons could be considered with the help of
perturbation theory. Indeed, let us consider two electron - hole pairs
(e1,h1) and (e2,h2) bound in excitons. One could expect that the Coulomb
interaction within pairs (i.e., the interaction between e1 and h1 and the
interaction between e2 and h2) has to be taken into account exactly because
it provides bound states, while the Coulomb interaction between particles
belonging to different excitons (e.g., the interaction between e1 and h2 and
the interaction between e2 and h1) can be considered as a perturbation.
However, due to electron-electron and hole-hole exchange it is impossible to
say if bound pairs are really (e1,h1) and (e2,h2) or (e1,h2) and (e2,h1).
For this reason it is not clear what part of the Coulomb interaction is the
interaction between excitons and can be considered as a perturbation.

A few approaches have been developed which formally related the gas of
electron-hole pairs to a Bose gas and avoided such difficulty.
Hanamura\cite{Hanamura70} made use of the Usui transformation, \cite{Usui57}
that makes a correspondence between the space of fermion pairs and a
"hypothetical" boson space. The transformed Hamiltonian still contained Fermi
operators. To eliminate them the commutators of this Hamiltonian with Bose
operators averaged over the ground state of the fermion system were declared
to be equal to the commutators of the target boson Hamiltonian with Bose
operators. The harmonic part of the resulting boson Hamiltonian immediately
led to the known exciton states. Thus, it was natural to assume that the
anharmonic part describes exciton-exciton interaction. Haug and
Schmitt-Rink\cite{Haug84} (see also Ref.\onlinecite{Tejedor96}) modified
this approach introducing creation and annihilation operators as linear
combinations of pairs of Fermi operators. Commutation relations of new
operators are different from those of Bose operators by terms proportional
to the density of bosons. The coefficients of the linear combination were
found from a variation principle and appeared to be single-exciton wave
functions corrected by interactions between bosons. Stolz {\it et
al}.\cite{stolz81} used another approach where they represented wave
functions of the electron-hole plasma as linear combinations of products of
single-exciton wave functions. Then again a variational principle was used
to find the single-exciton wave function. In all these works the
exciton-exciton interaction was calculated in the leading order in the
exciton density.

In this paper we develop a theory of exciton-exciton interaction in quantum
wells which is free of a formal definition of exciton creation and
annihilation operators. We do not use a variational method, the accuracy of
which is difficult to control. The physical basis to our approach is that we
consider a system of equal number of electrons and holes at low enough
temperature and low density, when all the particles are bound in excitons.
If the density is not small then excitons overlap and the Coulomb
interaction between particles of different excitons is of the same order as
the electron-hole interaction within one exciton. In such a case the
identity of excitons is lost and the electron-hole plasma can hardly be
considered as gas of excitons. If the other limitation, i.e., small kinetic
energy of excitons is not met then excitons can be ionized and the system
can not be described as the gas of excitons only. Experimentally small
exciton kinetic energy is achieved by resonant excitation at lattice
temperature much lower than the binding energy
\cite{Snoke,Jeune98,Shah,Baylac95,Amand97}.

Technically our approach resembles that of Stolz {\it et al}.\cite{stolz81}.
However, we don't rely on a variational principle but use a systematic
expansion in small exciton density. Finally, we arrive at an expression for
the second quantized exciton Hamiltonian with coefficients that are
expressed in terms of the free single-exciton wave function. This function,
and so the coefficients, strongly depend on the geometry of the
heterostructure.

We use the exciton Hamiltonian for the calculation of the blue shift of the
exciton luminescence line and the energy splitting between excitons with
different spin projection. These phenomena are described by Hamiltonian
matrix elements which are diagonal with respect to occupation numbers of
exciton states. Off-diagonal matrix elements of the Hamiltonian describe
exciton-exciton scattering. We use them to write down the Bolzmann equation
for the excitons and estimate the relaxation time in the exciton gas. For
numerical calculation of the coefficients of the Hamiltonian we use the
variational single-exciton wave function that we suggested earlier
\cite{smadar}.

We are not trying to reach an exact numerical matching of our theory with
experiments. This would require an accurate calculation of the
single-exciton wave function for different geometries. Our goal is rather to
explain semi-quantitatively (i.e., with an accuracy better than the order of
magnitude) as many experiments as possible. In other words, we are trying to
show that our theory is able to describe all so far detected phenomena in
the exciton gas of a small density which are related to exciton-exciton
scattering.

The structure of the paper reflects the specifics of the problem. We
consider electrons and holes confined in quantum wells. Their motion in the
growth direction is strongly quantized (we assume that only the ground
electron and hole states are occupied) and the problem is essentially
two-dimensional. So, in the next section we reduce the three-dimensional
(3D) Hamiltonian of electrons and holes to a 2D one. In Sec. \ref{deh} we
present the main assumptions of our approach and give the derivation of the
exciton Hamiltonian. Some cumbersome details of the derivation are
transferred to appendices. In Sec. \ref{sec:mfa} we calculate the blue shift
of the exciton luminescence line and the energy splitting of excitons with
opposite spins. In Sec. \ref{esr} we estimate the exciton-exciton relaxation
time with the help of the Bolzmann equation. We discuss our results and
compare them to experiments in Sec. \ref{sec:dis}. Our conclusions are
presented in Sec. \ref{con}.

\section{The Hamiltonian of 2D electron - hole plasma}
\label{Hehg}

Due to a strong quantization of the electron and hole motion in the growth
direction the dynamics of electrons and holes is essentially
two-dimensional. For this reason excitons in quantum wells sometimes are
considered as purely 2D\cite{Schmitt-Rink,Tejedor96,Ciuti98}. In this
approximation, however, it is impossible to describe effects of the
geometrical parameters of quantum wells on the exciton binding energy and
their interaction. So we use a more realistic model which takes into account
the geometry of the heterostructure.

In quantum wells, light and heavy holes are split off in energy. We consider
only heavy holes, assuming that the splitting is much larger than kinetic
energies of all involved particles and the interaction between them.
Therefore, light hole states are not occupied and the Hamiltonian of the
plasma of $N$ electrons and $N$ heavy holes in quantum wells is
\begin{eqnarray}
{\cal H}_{3D} & = & \sum_{j}
    \left[T_{ej} + T_{hj} + U_{0e}(z_{ej}) + U_{0h}(z_{hj})\right]
\label{eq:Hehg.1} \\ & + &
    \sum_{ij} U_{eh}(|\vec{r}_{ei} - \vec{r}_{hj}|, z_{ei}, z_{hj}) +
    {1 \over 2} \sum_{ij}   \left[
    U_{ee}(|\vec{r}_{ei} - \vec{r}_{ej}|, z_{ei}, z_{ej}) +
    U_{hh}(|\vec{r}_{hi} - \vec{r}_{hj}|, z_{hi}, z_{hj})
                            \right] \ .
\nonumber
\end{eqnarray}
Here $\vec{r}_{e}$ and $z_{e}$ are electron coordinates in the quantum well
plane and in the growth direction respectively, $\vec{r}_{h}$ and $z_{h}$ is
the same for holes,
\begin{mathletters}
\begin{eqnarray}
&&  T_{e}= -{{\hbar^2 \nabla_{e}^2}\over{2 m_e}}
        -{{\hbar^2}\over{2 m_e}}
    {{\partial^2}\over{\partial z_{e}^2}} \ ,
\label{eq:Hehg.2a}\\
&&  T_{h} =-{{\hbar^2 \nabla_{h}^2}\over{2 m_\parallel}}
    -{{\hbar^2}\over{2 m_\perp}}
        {{\partial^2}\over{\partial z_{h}^2}} \ ,
\label{eq:Hehg.2b}
\end{eqnarray}
\label{eq:Hehg.2}
\end{mathletters}
are the operators of the electron and hole kinetic energy, $m_{e}$ is the
electron mass, $m_{\parallel}$ and $m_{\perp}$ are in-plane and
perpendicular hole effective masses,  $\nabla_{e}$ and $\nabla_{h}$ are the
derivatives with respect to in-plane coordinates of electron and hole,
$U_{0e}(z_{e})$ and $U_{0h}(z_{h})$ are the heterostructure potentials that
confine electrons and holes in quantum wells, and $U_{eh}$, $U_{ee}$, and
$U_{hh}$ are the electron-hole, electron-electron, and hole-hole Coulomb
interaction energies respectively. If the difference between dielectric
constants in wells and barriers is negligible then
\begin{mathletters}
\begin{eqnarray}
&& U_{eh}(|\vec{r}_{1} - \vec{r}_{2}|, z_{1}, z_{2})
    = - {{e^2}\over{\kappa \sqrt{{(z_{1}-z_{2})}^2 +
    {(\vec{r}_{1}-\vec{r}_{2})}^2}}} \ ,
\label{eq:Hehg.3a} \\
&& U_{ee}(|\vec{r}_{1} - \vec{r}_{2}|, z_{1}, z_{2}) =
    U_{hh}(|\vec{r}_{1} - \vec{r}_{2}|, z_{1}, z_{2})
    ={{e^2}\over
        {\kappa \sqrt{{(z_{1}-z_{2})}^2 +
    {(\vec{r}_{1}-\vec{r}_{2})}^2}}} \ ,
\label{eq:Hehg.3b}
\end{eqnarray}
\label{eq:Hehg.3}
\end{mathletters}
where $\kappa$ is the dielectric constant.

The method that we develop can be applied to a single quantum well or to
coupled quantum wells with all possible well and barrier widths. In this
paper we use it only for coupled quantum wells where electrons and holes are
spatially separated. We neglect the electron-hole exchange\cite{Bir,sham}
which is very small in general \cite{Blackwood,Glasberg} and is further
reduced in coupled quantum wells due to the smaller electron-hole wave
function overlap. We also neglect the reduced symmetry of quantum well -
barrier interfaces since this effect is very small.\cite{aleiner} As a
result, Hamiltonian (\ref{eq:Hehg.1}) does not depend on electron and hole
spins.

The Hamiltonian (\ref{eq:Hehg.1}) contains an information about particle
confinement in quantum wells which makes the study of in-plane states of the
particles more difficult. Fortunately, in the most interesting cases the
in-plane motion can be separated and the problem is simplified. Typically
the interaction energy between the lowest level in a quantum well and the
first excited one (a few tens of meV or even larger) is much larger than
Coulomb interaction between particles (a few meV). For this reason at low
enough temperature particles are confined at the lowest quantization level.
A distortion of the wave functions caused by an admixture with the next
level due to Coulomb interaction between the particles can be neglected. In
such a case $z$-dependence of electron and hole wave functions is described
by $\zeta_{e}(z)$ and $\zeta_{h}(z)$ respectively, which satisfy the
equations
\begin{mathletters}
\begin{eqnarray}
&& -{{\hbar^2}\over{2 m_e}} \ {{\partial^2}\zeta_{e}\over{\partial z^2}} +
    U_{0e}(z)\zeta_{e} = E_{0e}\zeta_{e} \ ,
\label{eq:Hehg.4a} \\
&& -{{\hbar^2}\over{2 m_\perp}} \
    {{\partial^2}\zeta_{h}\over{\partial z^2}} + U_{0h}(z)\zeta_{h} =
    E_{0h}\zeta_{h} \ .
\label{eq:Hehg.4b}
\end{eqnarray}
\label{eq:Hehg.4}
\end{mathletters}
Here $E_{0e}$ and $E_{0h}$ are the electron and hole confinement energies
respectively. The wave function of the gas is a product of all single
particles wave functions $\zeta_{e}$ and $\zeta_{h}$ and a many-particle
wave function describing the in-plane state. The Hamiltonian that controls
the in-plane dynamics of the gas is obtained from the Hamiltonian,
(\ref{eq:Hehg.1}), by averaging with the functions $\zeta_{e}$ and
$\zeta_{h}$, and has the form
\begin{eqnarray}
{\cal H} & = & \sum_{j}
        \left(
    - {{\hbar^2 \nabla_{ej}^2}\over{2 m_e}}
    - {{\hbar^2 \nabla_{hj}^2}\over{2 m_\parallel}}
        \right)
\nonumber \\
&&   + \sum_{ij} u_{eh}(|\vec{r}_{ei} - \vec{r}_{hj}|) +
    {1 \over 2} \sum_{ij}   \left[
    u_{ee}(|\vec{r}_{ei} - \vec{r}_{ej}|) +
    u_{hh}(|\vec{r}_{hi} - \vec{r}_{hj}|)
                            \right] \ ,
\label{eq:Hehg.5}
\end{eqnarray}
where
\begin{equation}
u_{ij}(r) = \int U_{ij}(r, z_{1}, z_{2})
    \zeta_i^2(z_1) \zeta_j^2(z_2) dz_1 dz_2 \ .
\label{eq:Hehg.6}
\end{equation}
All details of the derivation of 2D Hamiltonian, (\ref{eq:Hehg.5}), from 3D
one, (\ref{eq:Hehg.1}), in the case of two particles are given in
Ref.\onlinecite{smadar}. The derivation of many particle Hamiltonian can be
done in the same way.

Thus to study electron - hole gas confined in quantum wells it is enough to
consider only a 2D problem described by Hamiltonian, (\ref{eq:Hehg.5}).

\section{Derivation of the exciton Hamiltonian}
\label{deh}

In this section we derive the 2D Hamiltonian of exciton gas starting from the
2D Hamiltonian of electrons and holes, Eq. (\ref{eq:Hehg.5}). We begin this
derivation with a discussion of the necessary conditions for considering the
electron - hole gas as exciton gas.

Not in any state the electron - hole gas can be represented as a gas of
excitons. To make this possible, two conditions are necessary. The first one
is a small enough concentration of electrons and holes, $n$. Under this
condition all electrons and holes can be bound in excitons and the excitons
are far from each other. That is
\begin{equation}
na^{2} \ll 1 \ ,
\label{eq:deh.1}
\end{equation}
where $a$ is the exciton radius. If this condition is not met then excitons
overlap and the Coulomb interaction between electrons and holes of different
excitons becomes of the order of the interaction within one exciton. In such
a case it is impossible to identify excitons, and the electron - hole plasma
hardly can be described as a gas of excitons.

The other condition is that typical exciton kinetic energy is much smaller
than the absolute value of the exciton binding energy, $\epsilon_b$,
\begin{equation}
{\hbar^{2}K^{2} \over 2M} \ll \epsilon_{b} \ .
\label{eq:deh.2}
\end{equation}
Here $\vec{K}$ and $M$ are the exciton wave vector and the exciton mass
respectively. If this condition is not met then as a result of collisions
between excitons they can be excited from the ground state or even be
ionized into free electrons and holes. Then it is necessary to consider
excitons interacting with each other and with free electrons and
holes.\cite{Haug76} We are not going to consider this case. Strictly
speaking, the accurate condition for ground state excitons contains the
energy separation between the ground state and the first excited state of
exciton in the right hand side of Eq. (\ref{eq:deh.2}). But this difference
is of the order of $\epsilon_b$ and we can use Eq. (\ref{eq:deh.2}) as it is.

To derive the exciton Hamiltonian we construct a basis for the space of all
electron - hole states from products of single exciton wave functions. Under
the conditions (\ref{eq:deh.1}) and (\ref{eq:deh.2}) we expect that only the
ground state exciton wave function is important. So of all matrix elements of
Hamiltonian (\ref{eq:Hehg.5}) only the elements between the states
constructed of these functions should be kept. Such a program in an accurate
form is carried out in Sec. \ref{sec:deh.rh}, where we show how to calculate
all matrix elements of the exciton Hamiltonian. Due to inequality
(\ref{eq:deh.1}) some of these matrix elements are small. Namely, matrix
elements describing triple and higher order interaction between excitons are
proportional to higher power of the exciton concentration than matrix
elements describing pair interaction, and we neglect them. For the
calculation of the pair interaction matrix elements it is enough to consider
the system of only two excitons. This calculation is performed in Sec.
\ref{sec:deh.2eh}. Finally, in Sec. \ref{deh.heg}, we use these matrix
elements to write down the Hamiltonian of the exciton gas in the second
quantized form.

\subsection{Reduction of the electron - hole Hamiltonian to the exciton
Hamiltonian}
\label{sec:deh.rh}

\subsubsection{Exciton basis}

The starting point of our derivation is the Hamiltonian of many electrons
and holes, (\ref{eq:Hehg.5}). The general idea of the derivation is that we
reduce the space of all electron-hole states to the subspace of ground state
excitons states only, and project the Hamiltonian to this subspace. The
first step in this direction is the construction of a basis in which this
subspace can be separated.

We construct such a basis of symmetrized products of single exciton wave
functions. Single exciton wave functions are eigenfunctions of the
Hamiltonian
\begin{equation}
{\cal H}_{1eh} = - {{\hbar^2 \nabla_{e}^2}\over{2 m_e}}
    - {{\hbar^2 \nabla_{h}^2}\over{2 m_\parallel}} +
    u_{eh}(|\vec{r}_{e} - \vec{r}_{h}|) \ .
\label{eq:deh.rh.3}
\end{equation}
Since the Hamiltonian (\ref{eq:deh.rh.3}) is independent on electron and
hole spins, a single exciton wave function can be written as
\begin{equation}
\Psi_{\vec{K}\alpha,s}
    (\vec{r}_{e},\sigma_{e}; \vec{r}_h,\sigma_{h}) =
    g_{s}(\sigma) \psi_{\vec{K}\alpha}(\vec{r}_{e}; \vec{r}_{h}) \ .
\label{eq:deh.rh.4}
\end{equation}
Here, $\psi_{\vec{K}\alpha}(\vec{r}_{e}; \vec{r}_{h})$ is the eigenfunction
of Hamiltonian, (\ref{eq:deh.rh.3}), and the spin wave function,
\begin{equation}
g_{s}(\sigma) = \delta_{s,\sigma} \ ,
\label{eq:deh.rh.5}
\end{equation}
can be represented as the product of the electron and hole spin functions,
$g_{s}(\sigma)=g_{s_e}(\sigma_{e})g_{s_h}(\sigma_{h})$. The projection of
electron and hole spins to $z$-direction can take the values $s_{e}=\pm1/2$
and $s_{h}=\pm3/2$ respectively. The exciton spin projection to the same
direction is $s=s_{e}+s_{h}$ and it can take the values $\pm1,\pm2$. The
spin variables, $\sigma=\sigma_{e}+\sigma_{h}$, take the same values. There
is one to one correspondence between the set of electron and hole spins and
the exciton spin, i.e., each exciton spin corresponds to a single
combination of electron and hole spins.

The in-plane exciton wave function is
\begin{equation}
\psi_{\vec{K}\alpha}(\vec{r}_e,\vec{r}_{h})
    ={1\over\sqrt{S}}e^{i\vec{K} \vec{R} }
    \phi_{\alpha}(|\vec{r}_e-\vec{r}_h|) \ .
\label{eq:deh.rh.6}
\end{equation}
Here, $\vec{R}=({m_{e}\vec{r}_e+m_{\parallel}\vec{r}_h})/M$ is the exciton
center of mass coordinate, $M=m_e+m_\parallel$ is the exciton mass, $S$ is
the normalization area, and the function, $\phi_{\alpha}(r)$ is an
eigenfunction of the Hamiltonian
\begin{equation}
H_{1eh} = - {{\hbar^2 \nabla^2}\over{2 \mu}} + u_{eh}(r) \ ,
\label{eq:deh.rh.7}
\end{equation}
where
\begin{equation}
{1 \over \mu} = {1 \over m_{e}} + {1 \over m_{\parallel}} \ ,
\label{eq:deh.rh.8}
\end{equation}
is the reduced mass.

The functions $\Psi_{\vec{K}\alpha,s}(\vec{r}_{e},\sigma_{e};
\vec{r}_h,\sigma_{h})$ form a complete basis for electron-hole pair states.
That means that a complete basis for the gas consisting of $N$ electrons and
$N$ holes can be constructed of these functions. This basis consists of
correctly symmetrized products of single exciton functions,\cite{stolz81}
\begin{equation}
\Phi_{\{\nu\}}(\vec{r}_{e1},\sigma_{e1};\vec{r}_{h1},\sigma_{h1};
    \dots,\vec{r}_{eN},\sigma_{eN};\vec{r}_{hN},\sigma_{hN}) =
    {1 \over N!} \sum (-1)^{P} \prod_{j=1}^{N}
    \Psi_{\nu_{j}}
    (\vec{r}_{ej_{1}},\sigma_{ej_{1}}; \vec{r}_{hj_{2}},\sigma_{hj_{2}}) \ .
\label{eq:deh.rh.9}
\end{equation}
Here $\nu$ stands for the set of quantum numbers $(\vec{K},\alpha,s)$,
$\{\nu\}=\nu_{1},\nu_{2},\dots,\nu_{N}$, the summation is carried out over
all transpositions of electrons and holes, $j_{1}$ and $j_{2}$, and $P$ is
the parity of a transposition. Basis (\ref{eq:deh.rh.9}) is very convenient
for our purpose because it easily allows us to separate the subspace of wave
functions containing only ground state excitons. This subspace contains
those $\Phi_{\{\nu\}}$ in which all $\alpha$ correspond to the ground state.
We enumerate this subspace as 1 and the subspace of functions containing at
least one excited exciton as 2.

It is necessary to note that the functions $\Phi_{\{\nu\}}$ are not
orthogonal, in spite of the orthogonality of the single exciton functions
$\Psi_{\nu}$.\cite{stolz81} Let us, for instance, consider the integral of
the product of two functions $\Phi_{\{\nu\}}$ and $\Phi_{\{\nu^{\prime}\}}$
which differ by only one of all the quantum numbers $\nu$, e.g., $\nu_{1}$
in the first function is replaced by $\nu_{1}^{\prime}\neq\nu_{1}$ in the
second one. Comparing these two wave functions we see that for each term in
the sum (\ref{eq:deh.rh.9}) of one wave function there are terms in the sum
of the other that differ by exchange of pairs of electrons and holes in such
a way that the identity of excitons is not violated [e.g.,
$\Psi_{\nu_{1}}(\vec{r}_{e1},\sigma_{e1};\vec{r}_{h1},\sigma_{h1})
\Psi_{\nu_{2}}(\vec{r}_{e2},\sigma_{e2};\vec{r}_{h2},\sigma_{h2})$ and
$\Psi_{\nu_{1}^{\prime}}(\vec{r}_{e2},\sigma_{e2};\vec{r}_{h2},\sigma_{h2})
\Psi_{\nu_{2}}(\vec{r}_{e1},\sigma_{e1};\vec{r}_{h1},\sigma_{h1})$]. The
integrals of the products of these terms equal zero because of the
orthogonality of single exciton wave functions. But there are also terms
that differ by a transposition of electrons or holes which violates the
exciton identity, e.g.,
$\Psi_{\nu_{1}}(\vec{r}_{e1},\sigma_{e1};\vec{r}_{h1},\sigma_{h1})
\Psi_{\nu_{2}}(\vec{r}_{e2},\sigma_{e2};\vec{r}_{h2},\sigma_{h2})$ and
$\Psi_{\nu_{1}^{\prime}}(\vec{r}_{e1},\sigma_{e1};\vec{r}_{h2},\sigma_{h2})
\Psi_{\nu_{2}}(\vec{r}_{e2},\sigma_{e2};\vec{r}_{h1},\sigma_{h1})$. The
integral of the product of these terms is nonzero. From the definition of
the single exciton wave functions, Eq. (\ref{eq:deh.rh.6}), we see that this
integral, contains the factor $a^{2}/S$.

If functions $\Phi_{\{\nu\}}$ and $\Phi_{\{\nu^{\prime}\}}$ differ by more
than just one value of $\nu$ or the transposition mixes more than a pair of
excitons, the integral of $\Phi_{\{\nu\}}\Phi_{\{\nu^{\prime}\}}$ contains
the factor of $a^{2}/S$ to a higher power. Eventually, in the calculation of
observable quantities each factor of $1/S$ is accompanied by a sum over
occupied states of the system which gives the factor of $N$. So that the
nonorthogonality of the basis is characterized by the parameter $na^{2}$.

The basis (\ref{eq:deh.rh.9}) is not normalized. The normalization integral
of any of the basis functions, $\int |\Phi_{\{\nu\}}|^2 \prod_{j} d^2r_{ej}
d^2r_{hj}$, contains integrals of the same type of exchange that leads to the
nonorthogonality. Thus this integral is different from unity by terms of the
order of $na^{2}$.

\subsubsection{Wave equation for excitons}

After the characterization of the basis (\ref{eq:deh.rh.9}) we begin the
derivation of the exciton Hamiltonian. We write down the Schr\"odinger
equation for the gas of $N$ electrons and $N$ holes, ${\cal H}\Psi=E\Psi$,
in the matrix form in basis (\ref{eq:deh.rh.9}). To do this we need to
represent the eigenfunction $\Psi$ as an expansion in the functions
$\Phi_{\{\nu\}}$, and to obtain equations for the coefficients of this
expansion. For this purpose we multiply the Schr\"odinger equation by the
functions complex conjugated to $\Phi_{\{\nu\}}$ and integrate over all
variables. To write down the result in the matrix form we introduce the
notation $\Psi_{1}$ for the column of the expansion coefficients of the basis
function belonging to subspace 1 (only ground state excitons) and the
notation $\Psi_{2}$ for the column of the expansion coefficients of the
basis function belonging to subspace 2 (including also excited states). Then
the matrix equation takes the form
\begin{mathletters}
\begin{eqnarray}
&&  ({\cal H}_{11} - {\cal N}_{11}E) \Psi_{1} +
    ({\cal H}_{12} - {\cal N}_{12}E) \Psi_{2} = 0 \ ,
\label{eq:deh.rh.10a} \\
&&  ({\cal H}_{21} - {\cal N}_{21}E) \Psi_{1} +
    ({\cal H}_{22} - {\cal N}_{22}E) \Psi_{2} = 0 \ ,
\label{eq:deh.rh.10b}
\end{eqnarray}
\label{eq:deh.rh.10}
\end{mathletters}
where ${\cal H}_{ij}$ are the matrices with elements
$\langle\Phi_{\{\nu\}}|{\cal H}|\Phi_{\{\nu^{\prime}\}}\rangle$ and ${\cal
N}_{ij}$ are matrices with elements
$\langle\Phi_{\{\nu\}}|\Phi_{\{\nu^{\prime}\}}\rangle$, where
$\Phi_{\{\nu\}}$ and $\Phi_{\{\nu^{\prime}\}}$ belong to $i$ and $j$
subspaces respectively. We use Eq. (\ref{eq:deh.rh.10b}) in order to express
$\Psi_{2}$ in $\Psi_{1}$, substitute the result in Eq. (\ref{eq:deh.rh.10a})
and come up with the equation
\begin{equation}
({\cal H}_{11} - {\cal N}_{11}E) \Psi_{1} -
    {\cal H}^{(exex)} \Psi_{1} = 0 \ ,
\label{eq:deh.rh.11}
\end{equation}
where the effect of excited states is described by
\begin{equation}
{\cal H}^{(exex)} = ({\cal H}_{12} - {\cal N}_{12}E)
    ({\cal H}_{22} - {\cal N}_{22}E)^{-1}
    ({\cal H}_{21} - {\cal N}_{21}E) \ .
\label{eq:deh.rh.12}
\end{equation}
Although Eq. (\ref{eq:deh.rh.11}) contains only $\Psi_{1}$ it does not have
the form of the Schr\"odinger equation because ${\cal H}^{(exex)}$ is a
nonlinear function of the energy $E$. The reason is that Eq.
(\ref{eq:deh.rh.11}) is equivalent to Eq. (\ref{eq:deh.rh.10}) and describes
the general situation, where electrons and holes can occupy excited states.
To describe a system where all electrons and holes are bound in ground state
excitons we make use of the small parameters (\ref{eq:deh.1}) and
(\ref{eq:deh.2}).

\subsubsection{Exciton Hamiltonian}

Here we start from Eqs. (\ref{eq:deh.rh.11}) and (\ref{eq:deh.rh.12}),
neglect small terms and obtain the Hamiltonian and the Schr\"odinger
equation for the exciton gas.

First we note that ${\cal N}_{11}$ and ${\cal H}_{11}$ contain terms of
different order in $a^{2}/S$. The estimate of both ${\cal N}_{11}$ and
${\cal H}_{11}$ can be done in the same way. The larger the overlap between
$\Phi_{\{\nu\}}$ and $\Phi_{\{\nu^{\prime}\}}$, the larger the matrix
element $\langle\Phi_{\{\nu\}}|{\cal H}|\Phi_{\{\nu^{\prime}\}}\rangle$ is.
As we explained above, matrix elements between states that differ by a
transposition of two electrons or two holes contain the factor $a^{2}/S$ and
describe two-exciton interaction in ${\cal H}_{11}$. Matrix elements between
the states that differ by a transposition of three electrons or three holes,
e.g. $\Psi_{\nu_{1}}(\vec{r}_{e1},\sigma_{e1};\vec{r}_{h1},\sigma_{h1})
\Psi_{\nu_{2}}(\vec{r}_{e2},\sigma_{e2};\vec{r}_{h2},\sigma_{h2})
\Psi_{\nu_{3}}(\vec{r}_{e3},\sigma_{e3};\vec{r}_{h3},\sigma_{h3})$ and
$\Psi_{\nu_{1}^{\prime}}(\vec{r}_{e1},\sigma_{e1};\vec{r}_{h2},\sigma_{h2})
\Psi_{\nu_{2}^{\prime}}(\vec{r}_{e2},\sigma_{e2};\vec{r}_{h3},\sigma_{h3})
\Psi_{\nu_{3}^{\prime}}(\vec{r}_{e3},\sigma_{e3};\vec{r}_{h1},\sigma_{h1})$,
contain $(a^{2}/S)^{2}$. In ${\cal H}_{11}$ they describe a triple exciton
interaction that is not reduced to pair interaction. We neglect such and
other high order terms both in ${\cal N}_{11}$ and ${\cal H}_{11}$.

Matrix elements of ${\cal N}_{12}$, and ${\cal N}_{21}$ are nonzero only
because of the nonorthogonality of basis functions (\ref{eq:deh.rh.9}). As
we showed above $\langle\Phi_{\{\nu\}}|\Phi_{\{\nu^{\prime}\}}\rangle
\approx a^{2}/S$ when $\nu \neq \nu^{\prime}$. The matrix elements of ${\cal
H}_{12}$ and ${\cal H}_{21}$ contain the same parameter. To prove this, let
us examine the largest matrix elements of ${\cal H}_{12}$. The maximal
overlap between subspace 1 and subspace 2 is achieved when the state from
subspace 2 is such that all electrons and holes are bound in ground state
excitons except from one pair which is bound in an excited exciton state.
From the orthogonality of single exciton wave functions (\ref{eq:deh.rh.4})
with different $\alpha$ it follows that there are only two kinds of non
vanishing terms. The first are the terms that describe Coulomb interaction
between different excitons,
\begin{eqnarray}
&& \int |\psi_{K,\alpha}(\vec{r}_{e1},\vec{r}_{h1})|^2
    |\psi_{K,\alpha^\prime}(\vec{r}_{e2},\vec{r}_{h2})|^2
    d\vec{r}_{e1} d\vec{r}_{h1}d\vec{r}_{e2}d\vec{r}_{h2}
\nonumber \\ && \hspace{1cm} \times
        \left[
    u_{ee}(\vec{r}_{e1}-\vec{r}_{e2}) + u_{hh}(\vec{r}_{h1}-\vec{r}_{h2}) +
    u_{eh}(\vec{r}_{e1}-\vec{r}_{h2}) + u_{eh}(\vec{r}_{e2}-\vec{r}_{h1})
        \right] .
\label{eq:deh.rh.13}
\end{eqnarray}
This integral converges at large distance, $R$, between the excitons
(dipole-dipole interaction falls off as $1/R^{3}$) and it is of the order of
$(a^{2}/S)\epsilon_{b}$. The second kind are the terms where the identity of
the exciton is violated. We gave an example for those terms when we
discussed the nonorthogonality and saw that they contain the small parameter
$a^2/S$. Other non-vanishing terms, e.g. for states with smaller overlap
between subspace 1 and subspace 2 contain higher powers of $a^{2}/S$.

From the above arguments we see that the matrix ${\cal H}^{(exex)}$ contains
the small parameter $a^{2}/S$ coming from ${\cal N}_{12}$, ${\cal N}_{21}$,
${\cal H}_{12}$ and ${\cal H}_{21}$. For this reason, in the leading order,
all other contributions to ${\cal H}^{(exex)}$ containing this small
parameter can be neglected. In particular, in the summation over
intermediate states in Eq. (\ref{eq:deh.rh.12}) we consider only diagonal
terms of $({\cal H}_{22}-{\cal N}_{22}E)^{-1}$, and in these terms neglect
the interaction between excitons. In other words, ${\cal H}_{22}-{\cal
N}_{22}E$ can be replaced with the diagonal matrix $E_{\mu}-E$, where
$E_{\mu}$ is the sum of the energies of $N$ excitons with at least one of
them excited or ionized.

From this expression the importance of inequality (\ref{eq:deh.2}) is
immediately seen. If this inequality is not satisfied then the energy of the
ground state excitons $E$ can be close to $E_{\mu}$ due to the high ground
state exciton kinetic energy. In such a case some matrix elements of ${\cal
H}^{(exex)}$ contain a small denominator and become anomalously large. When
condition (\ref{eq:deh.2}) is met the exciton kinetic energy can be
neglected and $E\approx-N\epsilon_{b}$. As a result, matrix elements of
${\cal H}^{(exex)}$ are
\begin{equation}
{\cal H}_{\nu\nu^{\prime}}^{(exex)} = \sum_{\mu}
    {({\cal H}_{12} + {\cal N}_{12}N\epsilon_{b})_{\nu\mu}
    ({\cal H}_{21} + {\cal N}_{21}N\epsilon_{b})_{\mu\nu^{\prime}}
    \over E_{\mu} + N\epsilon_{b}} \ .
\label{eq:deh.rh.14}
\end{equation}
We would like to add one more comment concerning the neglect of the
exciton-exciton interaction in the intermediate states. The radius of highly
excited exciton states can be of the order of the distance between different
excitons and the interaction between excitons in this case is of the order
of the electron-hole interaction within the same exciton. However, such
highly excited states are very close to states of free electrons and holes
where Coulomb interaction can be neglected compared to the kinetic energy.

Eq. (\ref{eq:deh.rh.11}) with ${\cal H}^{(exex)}$ determined by Eq.
(\ref{eq:deh.rh.14}) is now linear in the energy $E$, and different from
Schr\"odinger equation just by the non-diagonal matrix ${\cal N}_{11}$. The
transformation that reduces Eq. (\ref{eq:deh.rh.11}) to the regular
Schr\"odinger equation is equivalent to the introduction of an orthogonal
and normalized basis, $\tilde{\Phi}_{\nu}$,
\begin{equation} \Phi_{\nu} =
{\cal V}_{\nu\nu^{\prime}}\tilde{\Phi}_{\nu^{\prime}} \ ,
\label{eq:deh.rh.15}
\end{equation}
where the matrix ${\cal V}$ is not unitary. (Note that this matrix is defined
only in subspace 1). The corresponding transformation of the ${\cal N}$
matrix is
\begin{equation}
{\cal V}^{\dag}{\cal N}{\cal V} = I \ ,
\label{eq:deh.rh.16}
\end{equation}
where $I$ is the unit matrix. Since the difference between ${\cal N}$ and
the unit matrix is small, we can write
\begin{equation}
{\cal N} = I + {\cal A} \ ,
\label{eq:deh.rh.17}
\end{equation}
where ${\cal A}^{\dag}={\cal A}$ and ${\cal A}\propto a^{2}/S$. That is,
${\cal V}$ can be found from Eq. (\ref{eq:deh.rh.16}) with the help of the
perturbation theory, assuming that ${\cal V}=I+{\cal V}_{1}$, where ${\cal
V}_{1}\sim a^{2}/S$. It is necessary to note that the transformation
(\ref{eq:deh.rh.16}) is not unique because an orthogonal and normalized
basis can be chosen by many ways which differ by unitary transformations. So
we chose a simplest solution to Eq. (\ref{eq:deh.rh.16}) that produces
minimal modification of the basis constructed of single exciton wave
functions, namely, ${\cal V}_{1}^{\dag}={\cal V}_{1}$. This immediately gives
\begin{equation}
{\cal V} = 1 - {1 \over 2} \ {\cal A} \ .
\label{eq:deh.rh.18}
\end{equation}
After the transformation to the orthogonal basis, Eq. (\ref{eq:deh.rh.11})
takes the following form
\begin{equation}
    {\cal V}^{\dag}({\cal H}_{11}- {\cal H}^{(exex)} ){\cal V} \Psi_{1}
    =E \Psi_{1} \ .
\label{eq:deh.rh.19}
\end{equation}
The matrix ${\cal H}_{11}$ can be represented as
\begin{equation}
{\cal H}_{11} = {\cal H}_{0} + {\cal H}_{1} \ ,
\label{eq:deh.rh.20}
\end{equation}
where ${\cal H}_{0}$ describes free excitons and ${\cal H}_{1}$ describes
their interaction, ${\cal H}_{1}\propto a^{2}/S$. We neglect terms of the
second and high order in $a^2/S$, so Eq. (\ref{eq:deh.rh.19}) is reduced to
\begin{mathletters}
\begin{eqnarray}
&& H_{ex} \Psi_{1} = E\Psi_{1} \ ,
\label{eq:deh.rh.21a} \\
&& H_{ex} = {\cal H}_{11} - {1\over2} \
    ({\cal A}{\cal H}_{0} + {\cal H}_{0}{\cal A}) + {\cal H}^{(exex)} \ .
\label{eq:deh.rh.21b}
\end{eqnarray}
\label{eq:deh.rh.21}
\end{mathletters}
Eq. (\ref{eq:deh.rh.21a}) is the Schr\"odinger equation for the gas of $N$
excitons and Eq. (\ref{eq:deh.rh.21b}) is the expression for the exciton
Hamiltonian that includes exciton-exciton interaction.

The necessity of the transition to a new basis, Eq. (\ref{eq:deh.rh.15}),
means that exciton-exciton interaction changes single exciton wave functions.
Such a change appears also in other
approaches\cite{Hanamura70,Haug84,stolz81}.

In the leading order in $na^{2}$ only matrix elements describing two exciton
interaction should be kept in Eq. (\ref{eq:deh.rh.21}). To calculate them it
is enough to consider the Hamiltonian of only two excitons. This is the
subject of the next subsection.

\subsection{Two-exciton Hamiltonian}
\label{sec:deh.2eh}

The Hamiltonian of two electrons and two holes has the following form
\begin{eqnarray}
{\cal H}_{2eh} & = & - {{\hbar^2 \nabla_{e1}^2}\over{2 m_e}}
    - {{\hbar^2 \nabla_{h1}^2}\over{2 m_\parallel}}
    - {{\hbar^2 \nabla_{e2}^2}\over{2 m_e}}
    - {{\hbar^2 \nabla_{h2}^2}\over{2 m_\parallel}} +
    u\left(\vec{r}_{e1},\vec{r}_{h1},\vec{r}_{e2},\vec{r}_{h2}\right) \ .
\label{eq:deh.2eh.1}
\end{eqnarray}
where
\begin{eqnarray}
&&  u\left( \vec{r}_{e1},\vec{r}_{h1},\vec{r}_{e2},\vec{r}_{h2}\right)=
    u_{ee}(|\vec{r}_{e1}-\vec{r}_{e2}|)+u_{eh}(|\vec{r}_{e1}-\vec{r}_{h1}|)
\nonumber \\
&&  +u_{eh}(|\vec{r}_{e1}-\vec{r}_{h2}|)+u_{eh}(|\vec{r}_{e2}-\vec{r}_{h1}|)
    +u_{eh}(|\vec{r}_{e2}-\vec{r}_{h2}|)+u_{hh}(|\vec{r}_{h1}-\vec{r}_{h2}|)
    \ .
\label{eq:deh.2eh.2}
\end{eqnarray}

For the calculation of matrix elements of ${\cal H}_{11}$ and ${\cal
N}_{11}$ it is necessary to know only the wave function (\ref{eq:deh.rh.9})
of two ground state excitons. Explicitly, this function is\cite{Ciuti98}
\begin{eqnarray}
&& \Phi_{\vec{K}_{1},s_1;\vec{K}_{2},s_2}
        (\vec{r}_{e1}\sigma_{e1},\vec{r}_{h1}\sigma_{h1},
        \vec{r}_{e2}\sigma_{e2},\vec{r}_{h2}\sigma_{h2})={1\over 2}
\nonumber \\
&&  \times \left[\Psi_{\vec{K}_1, s_1}
        (\vec{r}_{e1},\sigma_{e1}; \vec{r}_{h1},\sigma_{h1})
    \Psi_{\vec{K}_2,s_2}
        (\vec{r}_{e2},\sigma_{e2}; \vec{r}_{h2},\sigma_{h2}) \right.
\nonumber \\
&&  -\Psi_{\vec{K}_1,s_1}
        (\vec{r}_{e2},\sigma_{e2}; \vec{r}_{h1},\sigma_{h1})
    \Psi_{\vec{K}_2,s_2}
        (\vec{r}_{e1},\sigma_{e1}; \vec{r}_{h2},\sigma_{h2})
\nonumber \\
&&  -\Psi_{\vec{K}_1,s_1}
        (\vec{r}_{e1},\sigma_{e1}; \vec{r}_{h2},\sigma_{h2})
    \Psi_{\vec{K}_2,s_2}
        (\vec{r}_{e2},\sigma_{e2}; \vec{r}_{h1},\sigma_{h1})
\nonumber \\
&&  \left. +\Psi_{\vec{K}_1,s_1}
        (\vec{r}_{e2},\sigma_{e2}; \vec{r}_{h2},\sigma_{h2})
    \Psi_{\vec{K}_2,s_2}
        (\vec{r}_{e1},\sigma_{e1}; \vec{r}_{h1},\sigma_{h1})\right] \ .
\label{eq:deh.2eh.4}
\end{eqnarray}
Hereafter the quantum number $\alpha$, characterizing an internal exciton
state,  will be omitted in the case of the ground state. The unity matrix
elements in the space of two ground state excitons are
\begin{equation}
I_{\vec{K}_{1}s_{1},\vec{K}_{2}s_{2};
        \vec{K}_{3}s_{3},\vec{K}_{4}s_{4}} =
    \delta_{s_{1}s_{3}} \delta_{s_{2}s_{4}}
    \delta_{\vec{K}_{1},\vec{K}_{3}}
    \delta_{\vec{K}_{2},\vec{K}_{4}} +
    \delta_{s_{1}s_{4}} \delta_{s_{2}s_{3}}
    \delta_{\vec{K}_{1},\vec{K}_{4}}\delta_{\vec{K}_{3},\vec{K}_{4}} \ .
\label{eq:deh.2eh.5}
\end{equation}
The two products of $\delta$-symbols appear because the states are symmetric
with respect to the transposition of single exciton quantum numbers.

The matrix ${\cal N}_{11}$ in the subspace of two excitons states is ${\cal
N}_{\vec{K}_{1}s_{1},\vec{K}_{2}s_{2};
    \vec{K}_{3}s_{3},\vec{K}_{4}s_{4}}=\langle
\Phi_{\vec{K}_{1},s_1;\vec{K}_{2},s_2}|\Phi_{\vec{K}_{3},s_3;\vec{K}_{4},s_4}
    \rangle$.
According to the two exciton wave functions definition, (\ref{eq:deh.2eh.4}),
Eq. (\ref{eq:deh.rh.17}) has now the form ${\cal
N}_{\vec{K}_{1}s_{1},\vec{K}_{2}s_{2};
    \vec{K}_{3}s_{3},\vec{K}_{4}s_{4}}=I_{\vec{K}_{1}s_{1},\vec{K}_{2}s_{2};
    \vec{K}_{3}s_{3},\vec{K}_{4}s_{4}}+
    {\cal A}_{\vec{K}_{1}s_{1},\vec{K}_{2}s_{2};
    \vec{K}_{3}s_{3},\vec{K}_{4}s_{4}}$,
where
\begin{eqnarray}
&& {\cal A}_{\vec{K}_{1}s_{1},\vec{K}_{2}s_{2};
        \vec{K}_{3}s_{3},\vec{K}_{4}s_{4}} =
    {1 \over S} \ \delta_{\vec{K}_{1}+\vec{K}_{2},\vec{K}_{3}+\vec{K}_{4}}
\nonumber \\ && \hspace{1cm} \times
        \left(
    \delta_{s_{1e}s_{4e}}\delta_{s_{2e}s_{3e}}
    \delta_{s_{1h}s_{3h}}\delta_{s_{2h}s_{4h}} +
    \delta_{s_{1e}s_{3e}}\delta_{s_{2e}s_{4e}}
    \delta_{s_{1h}s_{4h}}\delta_{s_{2h}s_{3h}}
        \right) A \ ,
\label{eq:deh.2eh.6}
\end{eqnarray}
$s_{je}$ and $s_{jh}$ are electron and hole spins respectively of the exciton
with the spin $s_{j}$, and
\begin{eqnarray}
&&  A = - \int \phi_{q}^{4} \ {d\vec{q} \over (2\pi)^{2}} \ ,
\label{eq:deh.2eh.7}
\end{eqnarray}
where
\begin{equation}
\phi_{q} = \int e^{-i\vec{q}\vec{r}} \phi(r) \ d\vec{r} \ ,
\label{eq:deh.2eh.8}
\end{equation}
is the Fourier transform of the wave function. The reduction of the overlap
integral to such a simple form is possible due to the small parameter, Eq.
(\ref{eq:deh.2}). Details of the calculation are given in Appendix
\ref{sec:cme.oi}.

For the calculation of the matrix elements of ${\cal H}_{11}$ it is
convenient to separate them into two terms,
\begin{eqnarray}
\left({\cal H}_{11}\right)_{\vec{K}_{1}s_{1},\vec{K}_{2}s_{2};
    \vec{K}_{3}s_{3},\vec{K}_{4}s_{4}} =
    H_{\vec{K}_{1}s_{1},\vec{K}_{2}s_{2};
    \vec{K}_{3}s_{3},\vec{K}_{4}s_{4}}^{(d)} +
    H_{\vec{K}_{1}s_{1},\vec{K}_{2}s_{2};
    \vec{K}_{3}s_{3},\vec{K}_{4}s_{4}}^{(x)} \ .
\label{eq:deh.2eh.9}
\end{eqnarray}
The first part, $H^{(d)}$, contains matrix elements between initial and
final states where excitons consist of the same particles. This part consist
of 8 integrals which have similar form. The grouping of terms which are
different only by the notation of integration and summation variables leads
to
\begin{eqnarray}
H_{\vec{K}_{1}s_{1},\vec{K}_{2}s_{2};
        \vec{K}_{3}s_{3},\vec{K}_{4}s_{4}}^{(d)} & = &
    \sum_{\sigma_{e1}\sigma_{e2}\sigma_{h1}\sigma_{h2}}
    \int d\vec{r}_{e1} d\vec{r}_{h1} d\vec{r}_{e2} d\vec{r}_{h2}
\label{eq:deh.2eh.10} \\ && \hspace{-4cm} \times
    \Psi_{\vec{K}_{1}s_{1}}^{\ast}
    (\vec{r}_{e1}\sigma_{e1},\vec{r}_{h1}\sigma_{h1})
    \Psi_{\vec{K}_{2}s_{2}}^{\ast}
    (\vec{r}_{e2}\sigma_{e2},\vec{r}_{h2}\sigma_{h2})
        {\cal H}_{2eh}
    \Psi_{\vec{K}_{3}s_{3}}
    (\vec{r}_{e1}\sigma_{e1},\vec{r}_{h1}\sigma_{h1})
    \Psi_{\vec{K}_{4}s_{4}}
    (\vec{r}_{e2}\sigma_{e2},\vec{r}_{h2}\sigma_{h2})
\nonumber \\ & + &
    \sum_{\sigma_{e1}\sigma_{e2}\sigma_{h1}\sigma_{h2}}
    \int d\vec{r}_{e1} d\vec{r}_{h1} d\vec{r}_{e2} d\vec{r}_{h2} \ .
\nonumber \\ && \hspace{-4cm} \times
    \Psi_{\vec{K}_{1}s_{1}}^{\ast}
    (\vec{r}_{e1}\sigma_{e1},\vec{r}_{h1}\sigma_{h1})
    \Psi_{\vec{K}_{2}s_{2}}^{\ast}
    (\vec{r}_{e2}\sigma_{e2},\vec{r}_{h2}\sigma_{h2})
        {\cal H}_{2eh}
    \Psi_{\vec{K}_{4}s_{4}}
    (\vec{r}_{e1}\sigma_{e1},\vec{r}_{h1}\sigma_{h1})
    \Psi_{\vec{K}_{3}s_{3}}
    (\vec{r}_{e2}\sigma_{e2},\vec{r}_{h2}\sigma_{h2}) .
\nonumber
\end{eqnarray}
The second part, $H^{(x)}$, contains matrix elements between initial and
final states where excitons consist of different particles. Again we have 8
integrals that can be grouped to the following form
\begin{eqnarray}
H_{\vec{K}_{1}s_{1},\vec{K}_{2}s_{2};
        \vec{K}_{3}s_{3},\vec{K}_{4}s_{4}}^{(x)} = & - &
    \sum_{\sigma_{e1}\sigma_{e2}\sigma_{h1}\sigma_{h2}}
    \int d\vec{r}_{e1} d\vec{r}_{h1} d\vec{r}_{e2} d\vec{r}_{h2}
\label{eq:deh.2eh.11} \\ && \hspace{-4cm} \times
    \Psi_{\vec{K}_{1}s_{1}}^{\ast}
    (\vec{r}_{e1}\sigma_{e1},\vec{r}_{h1}\sigma_{h1})
    \Psi_{\vec{K}_{2}s_{2}}^{\ast}
    (\vec{r}_{e2}\sigma_{e2},\vec{r}_{h2}\sigma_{h2})
        {\cal H}_{2eh}
    \Psi_{\vec{K}_{3}s_{3}}
    (\vec{r}_{e2}\sigma_{e2},\vec{r}_{h1}\sigma_{h1})
    \Psi_{\vec{K}_{4}s_{4}}
    (\vec{r}_{e1}\sigma_{e1},\vec{r}_{h2}\sigma_{h2})
\nonumber \\ & - &
    \sum_{\sigma_{e1}\sigma_{e2}\sigma_{h1}\sigma_{h2}}
    \int d\vec{r}_{e1} d\vec{r}_{h1} d\vec{r}_{e2} d\vec{r}_{h2} \ .
\nonumber \\ && \hspace{-4cm} \times
    \Psi_{\vec{K}_{1}s_{1}}^{\ast}
    (\vec{r}_{e1}\sigma_{e1},\vec{r}_{h1}\sigma_{h1})
    \Psi_{\vec{K}_{2}s_{2}}^{\ast}
    (\vec{r}_{e2}\sigma_{e2},\vec{r}_{h2}\sigma_{h2})
        {\cal H}_{2eh}
    \Psi_{\vec{K}_{4}s_{4}}
    (\vec{r}_{e1}\sigma_{e1},\vec{r}_{h2}\sigma_{h2})
    \Psi_{\vec{K}_{3}s_{3}}
    (\vec{r}_{e2}\sigma_{e2},\vec{r}_{h1}\sigma_{h1}) .
\nonumber
\end{eqnarray}

There is a confusion in the literature about the term "exchange" in the
description of interaction (see, however, Ref.\onlinecite{Ciuti98}, where
clear definitions are given). For elementary particles, the first term in
Eq. (\ref{eq:deh.2eh.10}) is usually called "Hartree interaction" and the
second is called "exchange interaction" while $H^{(x)}$ does not exist. In
the case of excitons $H^{(d)}$ is called "direct interaction" and $H^{(x)}$
is called "exchange". We stick to this last terminology.

In the direct part, (\ref{eq:deh.2eh.10}), there are terms of ${\cal
H}_{2eh}$ which do not mix between the two excitons, i.e., the kinetic
energy of all the particles and the potential energy between particles
belonging to the same exciton. These terms will result with the unity
operator (\ref{eq:deh.2eh.5}) multiplied by the single exciton energy. The
other parts of ${\cal H}_{2eh}$ are the Coulomb interaction between the two
excitons. The details of the calculation are given in Appendix
\ref{sec:cme.dir} and the result is
\begin{eqnarray}
H_{\vec{K}_{1}s_{1},\vec{K}_{2}s_{2},
        \vec{K}_{3}s_{3},\vec{K}_{4}s_{4}}^{(d)} & = &
        \left(
    \delta_{s_{1}s_{3}} \delta_{s_{2}s_{4}}
    \delta_{\vec{K}_{1},\vec{K}_{3}}\delta_{\vec{K}_{2},\vec{K}_{4}} +
    \delta_{s_{1}s_{4}} \delta_{s_{2}s_{3}}
    \delta_{\vec{K}_{1},\vec{K}_{4}}\delta_{\vec{K}_{3},\vec{K}_{4}}
        \right)
    \left(E_{K_{3}} + E_{K_{4}}\right)
\nonumber \\ && +
    {1 \over S} \ \delta_{\vec{K}_{1}+\vec{K}_{2},\vec{K}_{3}+\vec{K}_{4}}
    U_{\vec{K}_{1}s_{1},\vec{K}_{2}s_{2},
        \vec{K}_{3}s_{3},\vec{K}_{4}s_{4}}^{(d)} \ ,
\label{eq:deh.2eh.12}
\end{eqnarray}
where $E_{K}=-\epsilon_{b}+{\hbar^{2}K^{2}/2M}$ is the single exciton energy, and
\begin{equation}
U_{\vec{K}_{1}s_{1},\vec{K}_{2}s_{2},
        \vec{K}_{3}s_{3},\vec{K}_{4}s_{4}}^{(d)} =
        \delta_{s_{1}s_{3}} \delta_{s_{2}s_{4}}
        U^{(d)}(|\vec{K}_{1}-\vec{K}_{3}|) +
        \delta_{s_{1}s_{4}} \delta_{s_{2}s_{3}}
        U^{(d)}(|\vec{K}_{1}-\vec{K}_{4}|) \ .
\label{eq:deh.2eh.13}
\end{equation}
Here
\begin{eqnarray}
U^{(d)}(q) & = &
        u_{ee}(q) \left[\phi^{2}\right]_{qm_{h}/M}^{2} +
        u_{hh}(q) \left[\phi^{2}\right]_{qm_{e}/M}^{2} +
        2u_{eh}(q)
        \left[\phi^{2}\right]_{qm_{h}/M} \left[\phi^{2}\right]_{qm_{e}/M} \ ,
\label{eq:deh.2eh.17}
\end{eqnarray}
where $u_{ij}(q)$ is the Fourier
component of the Coulomb potential,
\begin{equation}
u_{ij}(q)=\int u_{ij}(r) e^{-i\vec{q}\vec{r}} d\vec{r} \ ,
\label{eq:deh.2eh.15}
\end{equation}
and $\left[\phi^{2}\right]_{q}$ is the Fourier transform of the
wave function squared, according to the general definition,
\begin{equation}
\left[\phi\phi_{\alpha}\right]_{q} = \int e^{-i\vec{q}\vec{r}}
    \phi(r)\phi_{\alpha}(r) d\vec{r} \ .
\label{eq:deh.2eh.16}
\end{equation}

In the exchange part of the Hamiltonian matrix element,
(\ref{eq:deh.2eh.11}), all the terms mix between the excitons since the
initial and final state consist of different particles and therefore
corresponds to different pairs of excitons. The single exciton energy $E_K$
is now multiplied by the factor $A$ which reflects the small overlap between
the initial and final states. The result of the calculation is
\begin{eqnarray}
H_{\vec{K}_{1}s_{1},\vec{K}_{2}s_{2},
        \vec{K}_{3}s_{3},\vec{K}_{4}s_{4}}^{(x)}  & = &
    {U^{(x)} - 4A\epsilon_{b} \over S} \
    \delta_{\vec{K}_{1}+\vec{K}_{2},\vec{K}_{3}+\vec{K}_{4}}
\nonumber \\ && \hspace{-2cm} \times
        \left(
    \delta_{s_{1e}s_{4e}}\delta_{s_{2e}s_{3e}}
    \delta_{s_{1h}s_{3h}}\delta_{s_{2h}s_{4h}}+
    \delta_{s_{1e}s_{3e}}\delta_{s_{2e}s_{4e}}
    \delta_{s_{1h}s_{4h}}\delta_{s_{2h}s_{3h}}
        \right) ,
\label{eq:deh.2eh.18}
\end{eqnarray}
where $A$ was defined in Eq. (\ref{eq:deh.2eh.7}) and
\begin{eqnarray}
U^{(x)} & = &
    {\hbar^{2} \over \mu}
    \int \phi_{q}^{4} \ q^{2} \ {d\vec{q} \over (2\pi)^{2}} -
    \int \left[u_{ee}(r) + u_{hh}(r)\right]
        \left[
    \int \phi_{q}^{2} e^{i\vec{q}\vec{r}} {d\vec{q} \over (2\pi)^{2}}
        \right]^{2} d\vec{r} \ .
\label{eq:deh.2eh.19}
\end{eqnarray}
Details of the calculation are given in Appendix \ref{sec:cme.exc}.

Apparently, the first term in $H_{\vec{K}_{1}s_{1},\vec{K}_{2}s_{2},
        \vec{K}_{3}s_{3},\vec{K}_{4}s_{4}}^{(d)}$, Eq. (\ref{eq:deh.2eh.12}),
describes free excitons. According to the notation of Eq.
(\ref{eq:deh.rh.20}) this is a matrix element of ${\cal H}_{0}$. In other
words,
\begin{eqnarray}
{\cal H}_{0 \vec{K}_{1}s_{1} \vec{K}_{2}s_{2},
        \vec{K}_{3}s_{3},\vec{K}_{4}s_{4}} =
    \left(\delta_{s_{1}s_{3}} \delta_{s_{2}s_{4}}
    \delta_{\vec{K}_1,\vec{K}_3} \delta_{\vec{K}_2,\vec{K}_4}+
    \delta_{s_{1}s_{4}} \delta_{s_{2}s_{3}}
    \delta_{\vec{K}_1,\vec{K}_4} \delta_{\vec{K}_2,\vec{K}_3}
    \right)(E_{\vec{K}_3} +E_{\vec{K}_4}) \ .
\label{eq:deh.2eh.20}
\end{eqnarray}
The second term in $H_{\vec{K}_{1}s_{1},\vec{K}_{2}s_{2},
        \vec{K}_{3}s_{3},\vec{K}_{4}s_{4}}^{(d)}$, as well as
$H_{\vec{K}_{1}s_{1},\vec{K}_{2}s_{2},
        \vec{K}_{3}s_{3},\vec{K}_{4}s_{4}}^{(x)}$, describe exciton-exciton
interaction and, according to the same notation, are parts of the matrix
element of ${\cal H}_{1}$.

Matrix elements ${\cal H}_{\nu\nu^{\prime}}^{(exex)}$ seem to be of higher
order in the small parameter, $a^{2}/S$, since it contains the product of
two off-diagonal matrix elements, $({\cal H}_{12} + {\cal
N}_{12}N\epsilon_{b})_{\nu\mu}
    ({\cal H}_{21} + {\cal N}_{21}N\epsilon_{b})_{\mu\nu^{\prime}}$, each of
which is proportional to $a^2/S$. However, the additional factor of $a^2/S$
is cancelled by the summation with respect to the momentum of intermediate
states in Eq. (\ref{eq:deh.rh.14}). Therefore the order of magnitude of
${\cal H}_{\nu\nu^{\prime}}^{(exex)}$ is the same as of other matrix
elements describing pair exciton-exciton interaction, namely,
$\epsilon_{b}a^{2}/S$. Similar contribution to the exciton energy appears
also in other approaches where it is usually referred to as screening
correction. \cite{Tejedor96,Haug84,Zimmermann76,Nozieres82} The calculation
of ${\cal H}_{\nu\nu^{\prime}}^{(exex)}$, is much more complicated because
they contain wave functions of excited exciton states. The details of the
calculation of ${\cal H}_{\nu\nu^{\prime}}^{(exex)}$ are given in Appendix
\ref{sec:cme.exex}. The result has the form
\begin{eqnarray}
&& H_{\vec{K}_{1}s_{1},\vec{K}_{2}s_{2},
        \vec{K}_{3}s_{3},\vec{K}_{4}s_{4}}^{(exex)} =
    {1 \over S} \
    \delta_{\vec{K}_{1} + \vec{K}_{2}, \vec{K}_{3} + \vec{K}_{4}}
\nonumber \\ && \hspace{1cm} \times
        \Big[
        \left(
    \delta_{s_{1}s_{3}}\delta_{s_{2}s_{4}} +
    \delta_{s_{1}s_{4}}\delta_{s_{2}s_{3}}
        \right)
    V_{1}
\nonumber \\ && \hspace{1.2cm} -
        \left(
    \delta_{s_{e1}s_{e3}}\delta_{s_{h1}s_{h4}}
    \delta_{s_{e2}s_{e4}}\delta_{s_{h2}s_{h3}} +
    \delta_{s_{e1}s_{e4}}\delta_{s_{h1}s_{h3}}
    \delta_{s_{e2}s_{e3}}\delta_{s_{h2}s_{h4}}
        \right)
    V_{2}
        \Big] \ ,
\label{eq:deh.2eh.21}
\end{eqnarray}
where
\begin{mathletters}
\begin{eqnarray}
V_{1} & = &
    2 \sum_{\alpha_{1},\alpha_{2}} \int
    {|D_{\alpha_{1}\alpha_{2}}^{(d)}(\vec{q})|^{2} +
    |D_{\alpha_{1}\alpha_{2}}^{(x)}(\vec{q})|^{2} \over
    E_{\alpha_{1}} + E_{\alpha_{2}} +
    \hbar^{2}q^{2}/M + 2\epsilon_{b}} \
    {d\vec{q} \over (2\pi)^{2}} \ ,
\label{eq:deh.2eh.22a} \\
V_{2} & = &
    2 \sum_{\alpha_{1},\alpha_{2}} \int
    {D_{\alpha_{1}\alpha_{2}}^{(d)}(\vec{q})
    D_{\alpha_{1}\alpha_{2}}^{(x)\ast}(\vec{q}) +
    D_{\alpha_{1}\alpha_{2}}^{(d)\ast}(\vec{q})
    D_{\alpha_{1}\alpha_{2}}^{(x)}(\vec{q})
    \over E_{\alpha_{1}} + E_{\alpha_{2}} +
    \hbar^{2}q^{2}/M + 2\epsilon_{b}} \
    {d\vec{q} \over (2\pi)^{2}} \ .
\label{eq:deh.2eh.22b}
\end{eqnarray}
\label{eq:deh.2eh.22}
\end{mathletters}
Here in the sums with respect to the internal exciton quantum number $\alpha$
at least one of $\alpha_{1}$ and $\alpha_{2}$ corresponds to an excited
state. The remaining matrix elements are
\begin{mathletters}
\begin{eqnarray}
D_{\alpha_{1}\alpha_{2}}^{(d)}(\vec{q}) & = &
    u_{ee}(q) \left[\phi\phi_{\alpha_{1}}\right]_{-m_{\parallel}\vec{q}/M}
    \left[\phi\phi_{\alpha_{2}}\right]_{m_{\parallel}\vec{q}/M} +
    u_{hh}(q) \left[\phi\phi_{\alpha_{1}}\right]_{m_{\parallel}\vec{q}/M}
    \left[\phi\phi_{\alpha_{2}}\right]_{-m_{\parallel}\vec{q}/M}
\label{eq:deh.2eh.23a} \\ && +
    u_{eh}(q) \left[\phi\phi_{\alpha_{1}}\right]_{-m_{\parallel}\vec{q}/M}
    \left[\phi\phi_{\alpha_{2}}\right]_{-m_{e}\vec{q}/M} +
    u_{eh}(q) \left[\phi\phi_{\alpha_{1}}\right]_{m_{\parallel}\vec{q}/M}
    \left[\phi\phi_{\alpha_{2}}\right]_{m_{e}\vec{q}/M} \ ,
\nonumber \\
D_{\alpha_{1}\alpha_{2}}^{(x)}(\vec{q}) & = &
    \int    \Big[
    \phi_{\alpha_{1},-\vec{k}_{2}+\vec{q}m_{\parallel}/M}
    \phi_{\alpha_{2},-\vec{k}_{1}-\vec{q}m_{\parallel}/M}
    u_{ee}(|\vec{k}_{1} - \vec{k}_{2} + \vec{q}|)
\label{eq:deh.2eh.23b} \\ && \hspace{1cm} +
    \phi_{\alpha_{1},-\vec{k}_{1}-\vec{q}m_{e}/M}
    \phi_{\alpha_{2},-\vec{k}_{2}+\vec{q}m_{e}/M}
    u_{hh}(|\vec{k}_{1} - \vec{k}_{2} + \vec{q}|)
            \Big]
    \phi_{k_{1}}\phi_{k_{2}}
    {d\vec{k}_{1}d\vec{k}_{2} \over (2\pi)^{4}}
\nonumber \\ & + &
    \int
    \phi_{\alpha_{1},\vec{k}_{1}-\vec{q}m_{e}/M}
    \phi_{\alpha_{2},\vec{k}_{2}-\vec{q}m_{\parallel}/M}
    u_{eh}(|\vec{k}_{1} - \vec{k}_{2}|)
        \left(
    \phi_{k_{2}}\phi_{|\vec{k}_{2}-\vec{q}|} +
    \phi_{k_{1}}\phi_{|\vec{k}_{1}-\vec{q}|}
        \right)
    {d\vec{k}_{1}d\vec{k}_{2} \over (2\pi)^{4}} .
\nonumber
\end{eqnarray}
\label{eq:deh.2eh.23}
\end{mathletters}

According to Eq. (\ref{eq:deh.rh.21b}), to complete the calculation of the
interaction Hamiltonian it is necessary to calculate ${\cal A}{\cal
H}_{0}+{\cal H}_{0}{\cal A}$. Both ${\cal
A}_{\nu_{1}\nu_{2},\nu_{3}\nu_{4}}$ and ${\cal
H}_{0\nu_{1}\nu_{2},\nu_{3}\nu_{4}}$ are symmetric with respect to
transpositions of the first and the second pair of subscripts, i.e.,
\begin{mathletters}
\begin{eqnarray}
&& {\cal A}_{\nu_{1}\nu_{2},\nu_{3}\nu_{4}} =
    {\cal A}_{\nu_{2}\nu_{1},\nu_{3}\nu_{4}} =
    {\cal A}_{\nu_{1}\nu_{2},\nu_{4}\nu_{3}} \ ,
\label{eq:deh.2eh.24a} \\
&& {\cal H}_{0\nu_{1}\nu_{2},\nu_{3}\nu_{4}} =
    {\cal H}_{0\nu_{2}\nu_{1},\nu_{3}\nu_{4}} =
    {\cal H}_{0\nu_{1}\nu_{2},\nu_{4}\nu_{3}} \ .
\label{eq:deh.2eh.24b}
\end{eqnarray}
\label{eq:deh.2eh.24}
\end{mathletters}
They are defined in the symmetric space, i.e., $(\nu_{1}\nu_{2})$ and
$(\nu_{2}\nu_{1})$ describe the same two-exciton state. Because of this, the
summation with respect to intermediate states in ${\cal
A}_{\nu_{1}\nu_{2},\nu_{3}\nu_{4}}{\cal H}_{0\nu_{3}\nu_{4},\nu_{5}\nu_{6}}$
has to be carried out over all different pairs of $\nu_{3}\nu_{4}$, but not
independently with respect to $\nu_{3}$ and $\nu_{4}$. Actually, due to
continuous spectrum of $\vec{K}$ the contribution of diagonal states, where
$\nu_{3}=\nu_{4}$, can be neglected and then the independent summation
respect to $\nu_{3}$ and $\nu_{4}$ is equivalent to double counting of each
pair $\nu_{3}\nu_{4}$. As a result it is possible to sum independently with
respect to $\nu_{3}$ and $\nu_{4}$ and then to divide the result by 2. This
leads to
\begin{eqnarray}
\left({\cal A H}_0+{\cal H}_0 {\cal
    A}\right)_{\vec{K}_{1}s_{1},\vec{K}_{2}s_{2},
        \vec{K}_{3}s_{3},\vec{K}_{4}s_{4}}^{(x)} & = &
    {A \over S} \
    \delta_{\vec{K}_{1}+\vec{K}_{2},\vec{K}_{3}+\vec{K}_{4}}
\label{eq:deh.2eh.25}\\ && \hspace{-2cm} \times
        \left(
    \delta_{s_{1e}s_{4e}}\delta_{s_{2e}s_{3e}}
    \delta_{s_{1h}s_{3h}}\delta_{s_{2h}s_{4h}} +
    \delta_{s_{1e}s_{3e}}\delta_{s_{2e}s_{4e}}
    \delta_{s_{1h}s_{4h}}\delta_{s_{2h}s_{3h}}
        \right)
\nonumber \\ && \hspace{-2cm} \times
    \left(E_{\vec{K}_{1}} + E_{\vec{K}_{2}}
    + E_{\vec{K}_{3}} + E_{\vec{K}_{4}}\right)
\nonumber \\ & \approx &
    - {4A\epsilon_{b} \over S} \
    \delta_{\vec{K}_{1}+\vec{K}_{2},\vec{K}_{3}+\vec{K}_{4}}
\nonumber \\ && \hspace{-2cm} \times
        \left(
    \delta_{s_{1e}s_{4e}}\delta_{s_{2e}s_{3e}}
    \delta_{s_{1h}s_{3h}}\delta_{s_{2h}s_{4h}} +
    \delta_{s_{1e}s_{3e}}\delta_{s_{2e}s_{4e}}
    \delta_{s_{1h}s_{4h}}\delta_{s_{2h}s_{3h}}
        \right) .
\nonumber
\end{eqnarray}
Here we used again the assumption of a small exciton kinetic energy,
(\ref{eq:deh.2}).

Finally, we can present matrix elements of the two-exciton Hamiltonian in the
form
\begin{equation}
{\cal H}_{2ex} = {\cal H}_{0\vec{K}_{1}s_{1},\vec{K}_{2}s_{2},
    \vec{K}_{3}s_{3},\vec{K}_{4}s_{4}} +
    {1 \over S} \left(
    U_{\vec{K}_{1}s_{1},\vec{K}_{2}s_{2},
    \vec{K}_{3}s_{3},\vec{K}_{4}s_{4}}^{ex-ex} +
    U_{\vec{K}_{1}s_{1},\vec{K}_{2}s_{2},
    \vec{K}_{4}s_{4},\vec{K}_{3}s_{3}}^{ex-ex}
                \right) \ ,
\label{eq:deh.2eh.26}
\end{equation}
where matrix elements of ${\cal H}_{0}$ are defined by
Eq. (\ref{eq:deh.2eh.20}) and matrix elements of $U^{ex-ex}$ are
\begin{eqnarray}
U_{\vec{K}_{1}s_{1},\vec{K}_{2}s_{2},
        \vec{K}_{3}s_{3},\vec{K}_{4}s_{4}}^{ex-ex} & = &
    \delta_{\vec{K}_{1}+\vec{K}_{2},\vec{K}_{3}+\vec{K}_{4}}
    \delta_{s_{1}s_{3}} \delta_{s_{2}s_{4}}
    [U^{(d)}(|\vec{K}_{1}-\vec{K}_{3}|) + V_{1}]
\nonumber \\ & + &
    \delta_{\vec{K}_{1}+\vec{K}_{2},\vec{K}_{3}+\vec{K}_{4}}
    \delta_{s_{1e}s_{4e}}\delta_{s_{2e}s_{3e}}
    \delta_{s_{1h}s_{3h}}\delta_{s_{2h}s_{4h}}
    (U^{(x)} - 2A\epsilon_{b} - V_{2}) \ .
\label{eq:deh.2eh.27}
\end{eqnarray}
In this form we still keep a trace of non-elementary nature of excitons in
the spin $\delta$-symbols. The corresponding expression in exciton spins is
quite cumbersome. We overcome this disadvantage in the many-exciton
Hamiltonian in the next subsection.

\subsection{Hamiltonian of exciton gas with pair interaction}
\label{deh.heg}

The matrix form of the Hamiltonian of the exciton gas (\ref{eq:deh.rh.21b})
is practically inconvenient. It is desirable to reduce it to the second
quantized form. There is a standard way to obtain the second quantized form
of the Hamiltonian of a gas of elementary bosons. To make use of this way we
show that the Hamiltonian of two excitons (\ref{eq:deh.2eh.26}) is
equivalent to a Hamiltonian of two elementary bosons. Then, keeping only pair
exciton interaction, we can immediately write down the Hamiltonian of $N$
excitons.

The Hamiltonian of two elementary bosons with masses $M$ and the interaction
$\hat{U}$ can be written as
\begin{equation}
H_{ex} = - 2 \epsilon_{b} -
    {\hbar^{2} \over 2M} \ \nabla_{1}^{2} -
    {\hbar^{2} \over 2M} \ \nabla_{2}^{2} + \hat{U} \ ,
\label{eq:deh.heg.1}
\end{equation}
where $\epsilon_b$ is the energy necessary to create a boson.

As a basis for wave functions we chose plane waves so that wave function of
two bosons can be written as
\begin{mathletters}
\begin{equation}
\Psi_{\vec{K}_{1}s_{1},\vec{K}_{2}s_{2}} =
    {1 \over \sqrt{2}S}
                \left[
        g_{s_{1}}(\sigma_{1}) g_{s_{2}}(\sigma_{2})
        e^{i(\vec{K}_{1}\vec{R}_{1} + \vec{K}_{2}\vec{R}_{2})} +
        g_{s_{1}}(\sigma_{2}) g_{s_{2}}(\sigma_{1})
        e^{i(\vec{K}_{1}\vec{R}_{2} + \vec{K}_{2}\vec{R}_{1})}
                \right] ,
\label{eq:deh.heg.2a}
\end{equation}
when at least one of the inequalities $s_{1}\neq s_{2}$, $\vec{K}_{1}\neq
\vec{K}_{2}$ is satisfied and
\begin{equation}
\Psi_{\vec{K}_{1}s_{1},\vec{K}_{1}s_{1}} = {1 \over S} \
        g_{s_{1}}(\sigma_{1}) g_{s_{1}}(\sigma_{2})
        e^{i\vec{K}_{1}(\vec{R}_{1} + \vec{R}_{2})} \ ,
\label{eq:deh.heg.2b}
\end{equation}
\label{eq:deh.heg.2}
\end{mathletters}
when both spins and wave vectors of the bosons are equal.

Matrix elements of the Hamiltonian, (\ref{eq:deh.heg.1}), between functions,
(\ref{eq:deh.heg.2a}), are identical to the matrix elements,
(\ref{eq:deh.2eh.26}), if we identify the matrix elements of $\hat{U}$
between wave functions, (\ref{eq:deh.heg.2a}),
$\tilde{U}_{\vec{K}_{1}s_{1},\vec{K}_{2}s_{2},
        \vec{K}_{3}s_{3},\vec{K}_{4}s_{4}}^{ex-ex}+
        \tilde{U}_{\vec{K}_{1}s_{1},\vec{K}_{2}s_{2},
        \vec{K}_{4}s_{4},\vec{K}_{3}s_{3}}^{ex-ex}$ with
$U_{\vec{K}_{1}s_{1},\vec{K}_{2}s_{2},
        \vec{K}_{3}s_{3},\vec{K}_{4}s_{4}}^{ex-ex}+
        U_{\vec{K}_{1}s_{1},\vec{K}_{2}s_{2},
        \vec{K}_{4}s_{4},\vec{K}_{3}s_{3}}^{ex-ex}$,
Eqs. (\ref{eq:deh.2eh.26})-(\ref{eq:deh.2eh.27}). This last identification
is the definition of $\hat{U}$. The nontrivial spin structure of these
matrix elements means that the operator $\hat{U}$ is spin-dependent. [One
should remember that $s_{j}=s_{je}+s_{jh}$,
$\sigma_{j}=\sigma_{je}+\sigma_{jh}$, and
$g_{s}(\sigma)=g_{s_{e}}(\sigma_{e})g_{s_{h}}(\sigma_{h})$.] Due to the
continuity of $\vec{K}$, the contribution of matrix elements with
$\vec{K}_{1}=\vec{K}_{2}$ and with $\vec{K}_{3}=\vec{K}_{4}$ to all effects
is negligible and we don't consider them.

Now we have two-exciton Hamiltonian with known matrix elements of the pair
interaction operator. The many excitons Hamiltonian with pair interaction in
the second quantized form is usually written with the help of a
non-symmetrized matrix element, \cite{Landau}
\begin{eqnarray}
\tilde{U}_{\vec{K}_{1}s_{1},\vec{K}_{2}s_{2},
    \vec{K}_{4}s_{4},\vec{K}_{3}s_{3}}^{ex-ex} & = &
    {1 \over S} \sum_{\sigma_{1}\sigma_{2}}
    g_{s_{1}}(\sigma_{1})g_{s_{2}}(\sigma_{2})
\nonumber \\ && \times
    \int d\vec{R}_{1} d\vec{R}_{2} \
    e^{-i(\vec{K}_{1}\vec{R}_{1} + \vec{K}_{2}\vec{R}_{2})}
    \hat{U}_{ex-ex}
    e^{i(\vec{K}_{3}\vec{R}_{1} + \vec{K}_{4}\vec{R}_{2})} \
    g_{s_{3}}(\sigma_{1})g_{s_{4}}(\sigma_{2}) \ .
\label{eq:deh.heg.3}
\end{eqnarray}
Due to commutation relations of Bose operators it is possible to write down
this Hamiltonian also with the symmetrized matrix element,
$(1/2)(\tilde{U}_{\vec{K}_{1}s_{1},\vec{K}_{2}s_{2},
        \vec{K}_{3}s_{3},\vec{K}_{4}s_{4}}^{ex-ex}+
        \tilde{U}_{\vec{K}_{1}s_{1},\vec{K}_{2}s_{2},
        \vec{K}_{4}s_{4},\vec{K}_{3}s_{3}}^{ex-ex})$. That is, an additional
factor of 1/2 appears in the interaction term of the many exciton
Hamiltonian
\begin{eqnarray}
&&  H_{ex}= \sum_{K,s} E_K c^\dag_{K,s} c_{K,s}+
\label{eq:deh.heg.4} \\
&&  + {1\over {4 S}} \sum_{\vec{K}_{1},\vec{K}_{2},\vec{K}_{3},\vec{K}_{4}
    \atop s_{1},s_{2},s_{3},s_{4}}
        \left\{
        \left[
    \left(U^{(d)}(|K_1-K_3|) + V_{1}\right)
    \delta_{s_1,s_3} \delta_{s_2,s_4} +
    \left(U^{(d)}(|K_1-K_4|) + V_{1}\right)
    \delta_{s_1,s_4} \delta_{s_2,s_3}
        \right]
        \right.
\nonumber \\
&&      \left.  + V_{x}
        \left(
    \delta_{s_{e1},s_{e4}} \delta_{s_{e2},s_{e3}}
    \delta_{s_{h1},s_{h3}} \delta_{s_{h2},s_{h4}} +
    \delta_{s_{e1},s_{e3}} \delta_{s_{e2},s_{e4}}
    \delta_{s_{h1},s_{h4}} \delta_{s_{h2},s_{h3}}
        \right)
        \right\}
\nonumber \\
&&  \times
    \delta_{K_1+K_2,K_3+K_4} c^\dag_{K_2,s_2}  c^\dag_{K_1,s_1}
    c_{K_3,s_3} c_{K_4,s_4} .
\nonumber
\end{eqnarray}
where $c_{\vec{K}s}$ and $c_{\vec{K}s}^{\dag}$ are exciton annihilation and
creation operators respectively, and
\begin{equation}
V_{x}=U^{(x)}-2 A \epsilon_b - V_{2} \ .
\label{eq:deh.heg.5}
\end{equation}

Now we make use of the one to one correspondence between electron and hole
spins and the exciton spin, in order to express the $\delta$-symbols
containing separately electron and hole spins, in $\delta$-symbols
containing exciton spin only. Details of this calculation are given in
Appendix \ref{xsr}, and the result is
\begin{eqnarray}
H_{ex} & = & \sum_{K,s} E_K c^\dag_{K,s} c_{K,s}
    + {1\over {2 S}} \sum_{\vec{K}_{1},\vec{K}_{2},q \atop s_{1},s_{2}}
    \left[U^{(d)}(q) + V_{1}\right]
    c^\dag_{K_2,s_2} c^\dag_{K_1,s_1} c_{K_1-q,s_1} c_{K_2+q,s_2}
\nonumber \\ & + &
    {V_{x} \over 4 S} \sum_{\vec{K}_{1},\vec{K}_{2},q}
    \Bigg[
        \sum_{s_{1}s_{2}}
        c_{\vec{K}_{1}s_{1}}^{\dag}c_{\vec{K}_{2}-s_{1}}^{\dag}
        c_{\vec{K}_{1}-\vec{q}s_{2}}c_{\vec{K}_{2}+\vec{q}-s_{2}}
\nonumber \\
&&      -4 \sum_{s}
        c_{\vec{K}_{1}s}^{\dag}c_{\vec{K}_{2}-s}^{\dag}
        c_{\vec{K}_{1}-\vec{q}s}c_{\vec{K}_{2}+\vec{q}-s} +
        2 \sum_{s_{1}s_{2}}
        c_{\vec{K}_{1}s_{1}}^{\dag}c_{\vec{K}_{2}s_{2}}^{\dag}
        c_{\vec{K}_{1}-\vec{q}s_{1}}c_{\vec{K}_{2}+\vec{q}s_{2}}
        \Bigg] \ .
\label{eq:deh.heg.6}
\end{eqnarray}

Hamiltonian, (\ref{eq:deh.heg.6}), is the main result of this section. In
the following sections we use this Hamiltonian to study the density
dependent luminescence line shift and exciton-exciton relaxation. To do this
we calculate numerically the matrix elements of the Hamiltonian for a number
of structures. The calculation of $V_{1}$ and $V_{2}$ is extremely difficult
because it involves all exciton excited states. Fortunately, for the most of
the structures that we consider $V_{1}$ and $V_{2}$ are numerically small
compared to other matrix elements. The reason behind this is that the wave
functions of excited states oscillate in the region where the wave function
of the ground state is smooth. Our estimates show that the value of ${\cal
H}_{\nu\nu^{\prime}}^{(exex)}$ is smaller then 10\% of $A \epsilon_b$ or
$U^{(x)}$. So in the following calculation $V_{1}$ and $V_{2}$ are neglected.

\section{Mean field approximation. Luminescence line shift}
\label{sec:mfa}

In the mean field approximation the exciton scattering is neglected and only
the shift of the exciton energy due to interaction is taken into account.
This means that only terms diagonal with respect to occupation number
$n_{\vec{K}s}=c_{\vec{K}s}^{\dag}c_{\vec{K}s}$ are kept in the Hamiltonian.
Then the energy of the system of $N$ excitons is
\begin{eqnarray}
E_N & = &\sum_{K,s} E_K n_{\vec{K}s} +
    {V_{b} \over 2 S}
    N^2 +  {V_{x} \over 4 S} \sum_{s}\left(N_s-N_{-s}\right)^2
\nonumber \\ &&  +
    {1\over {2 S}} \sum_{\vec{K}_{1},\vec{K}_{2},s}
    U^{(d)}(|\vec{K}_1-\vec{K}_2|)
    n_{\vec{K}_1,s} n_{\vec{K}_2,s}
\label{eq:mfa.1}
\end{eqnarray}
where $N_{s}=\sum_{K} n_{\vec{K},s}$ is the number of excitons with spin
$s$, $N=\sum_{s}N_{s}$, and
\begin{equation}
V_{b} = U^{(d)}(0) + V_{x} \ .
\label{eq:mfa.2}
\end{equation}

The recombination energy $\epsilon_{\vec{K},s}$ of an exciton with momentum
$\vec{K}$ and spin $s$ equals the change of the energy of the exciton system
when the occupation number $n_{\vec{K} s}$ decreases by one,
$\epsilon_{\vec{K},s}=E_{N}(n_{\vec{K}s})-E_{N}(n_{\vec{K}
s}-1)\approx\partial E_{N}/\partial n_{\vec{K}s}$. Neglecting the photon
wave vector we can put $\vec{K}=0$ and then Eq. (\ref{eq:mfa.1}) leads to
\begin{eqnarray}
\epsilon_{s} = - \epsilon_{b} + V_{b} n + V_{x} (n_s-n_{-s})
    + \int {d^2 K \over (2 \pi)^2} U^{(d)}(K) n_{\vec{K},s} \ .
\label{eq:mfa.3}
\end{eqnarray}
where $n_s=N_s/S$ is the density of excitons with spin $s$ and $n=N/S$ is the
total exciton density. The energy splitting between optically active
excitons with spins +1 and -1 is
\begin{equation}
\Delta E_{1,-1} = \epsilon_{1} - \epsilon_{-1} =
    2 V_{x} (n_{1} - n_{-1}) +
    \int {d^2 K \over (2 \pi)^2} U^{(d)}(K)
    (n_{\vec{K},1}-n_{\vec{K},-1}) \ .
\label{eq:mfa.4}
\end{equation}

Eqs. (\ref{eq:mfa.3}) and (\ref{eq:mfa.4}) are complicated to use due to the
integral term. To calculate it, the exciton concentration is not enough, it
is necessary to know the energy distribution of excitons. These expressions
can be simplified if the typical exciton wave vector is much smaller than
the typical wave vector of $U^{(d)}(K)$. The last one is characterized by the
wells width, $L$ and the width of the barrier, $w$ (see Appendix
\ref{sec:fc}). The exciton wave vector is small in the case of resonant
pumping at low temperature $T$. Then it is of the order of the photon wave
vector $K_{ph}$ or of the thermal wave vector $K_{T}\sim\sqrt{2M T}/\hbar$.
So, if both $K_{ph}$ and $K_{T}$ are much smaller than $1/L$ and $1/w$ then
$U^{(d)}(K)$ in Eq. (\ref{eq:mfa.1}), (\ref{eq:mfa.3}) and (\ref{eq:mfa.4})
can be replaced with $U^{(d)}(0)$. As a result we have
\begin{eqnarray}
E_N = \sum_{K,s} E_K n_{\vec{K}s} +
    {V_{b} \over 2 S} N^2 +
    {V_{x} \over 4 S} \sum_{s}\left(N_s-N_{-s}\right)^2 +
    {U^{(d)}(0) \over {2 S}} \sum_{s} N_s^2 \ ,
\label{eq:mfa.5}
\end{eqnarray}
and
\begin{mathletters}
\begin{eqnarray}
&&  \epsilon_{s} = - \epsilon_{b} + V_{b} n + V_{x} (n_s-n_{-s})
    +  U^{(d)}(0) n_s \ ,
\label{eq:mfa.6a} \\
&&  \Delta E_{1,-1} = V_{es} (n_1-n_{-1}) \ ,
\label{eq:mfa.6b}
\end{eqnarray}
\label{eq:mfa.6}
\end{mathletters}
where $V_{es}=2 V_{x} + U^{(d)}(0)$. An expression similar to Eq.
(\ref{eq:mfa.6a}) was suggested phenomenologically by Amand {\it et
al}.\cite{Amand94}. This expression differs from Eq. (\ref{eq:mfa.6a}) by a
relation between the coefficients in two last terms.

From Eq. (\ref{eq:mfa.5}) we see that for a constant exciton number, $N$,
the ground state of the exciton gas can be paramagnetic or ferromagnetic,
depending on the sign of $V_{es}$. When $V_{es}$ is positive the minimal
total energy of the system is reached when the number of excitons with
opposite spins is equal, $N_{s}=N_{-s}$. For negative $V_{es}$, the system
reaches its minimal energy when the difference, $N_{s}-N_{-s}$, is maximal,
which corresponds to ferromagnetic phase. The possibility of these two
phases for condensed excitons has been pointed out by Fern\'{a}ndez-Rossier
and Tejedor. \cite{Tejedor97} From Eq. (\ref{eq:mfa.6b}) we see that the same
parameter, $V_{es}$, characterizes the spin energy splitting.

To make a quantitative comparison to experiments we evaluate the values of
the Hamiltonian matrix elements for coupled quantum wells where electrons
and holes are confined in different wells. Such a separation is usually
reached by an external electric field applied in the growth direction. In
the region of well widths that we consider the energy separation between the
ground state and the first excited state in the wells is much larger than
typical external potential drop across one well. This means that the
electron and hole ground state wave functions can be taken as
\begin{mathletters}
\begin{eqnarray}
\zeta_e(z_e) & = & \sqrt{2\over L_e} \sin{{\pi z_e \over L_e}} \ ,
    \hspace{1.5cm} -L_e<z_e<0 \ ,
\label{eq:mfa.7a} \\
\zeta_h(z_h) & = & \sqrt{2\over L_h} \sin{{\pi(z_h-w) \over L_h}} \ ,
    \hspace{0.5cm} w<z_h<w+L_h \ ,
\label{eq:mfa.7b}
\end{eqnarray}
\label{eq:mfa.7}
\end{mathletters}
where $L_{e}$ and $L_{h}$ and the widths of the electron and hole quantum
wells respectively and $w$ is the width of the barrier. The expressions for
the interaction energy $u_{ij}$ are cumbersome and their Fourier transforms
are presented in Appendix \ref{sec:fc}.

For the calculation of matrix elements we use the variational single-exciton
wave function that gives a very good approximation\cite{smadar},
\begin{equation}
\phi(r)={1\over \sqrt{2 \pi b (b+r_0)}}
    \exp\left\{-{{\sqrt{r^2+r_0^2}-r_0}\over 2 b}\right\} ,
\label{eq:mfa.8}
\end{equation}
where $b$ and $r_0$ are variation parameters that are found by minimizing of
the binding energy. The Fourier transform $\phi_{q}$ necessary for the
calculation of $U^{(d)}(q)$ and $V_{x}$ is given in Appendix \ref{sec:fc}.

For $U^{(d)}(0)$ it is possible to obtain a simple analytic expression
without making use of the exciton wave function (Appendix \ref{sec:fc}),
\begin{equation}
U^{(d)}(0) = {4\pi e^{2} \over \kappa}
    \left[w + 0.397 (L_{e} + L_{h})\right] .
\label{eq:mfa.9}
\end{equation}
We see that the direct interaction, $U^{(d)}(0)$, grows with the separation
between the wells, $d=w+(L_e+L_h)/2$. This behavior is easy to understand.
From the point of view of electrostatics, excitons resemble parallel dipoles
of the size $d$, and $U^{(d)}(0)$ is a the dipole - dipole interaction. This
interaction is a repulsion growing with the size of the dipoles. The
coefficient of the second term in the square brackets is close to 0.5, for
which $U^{(d)}(0)$ corresponds to the plate capacitor approximation
\cite{Butov,Yoshioka90,Zhu95}. Typical value for $U^{(d)}(0)$ for coupled
quantum wells where the separation between the wells is $d=100 $\AA \ is
about $1.5\times 10^{-10}$ meV$\cdot$ cm$^{2}$.

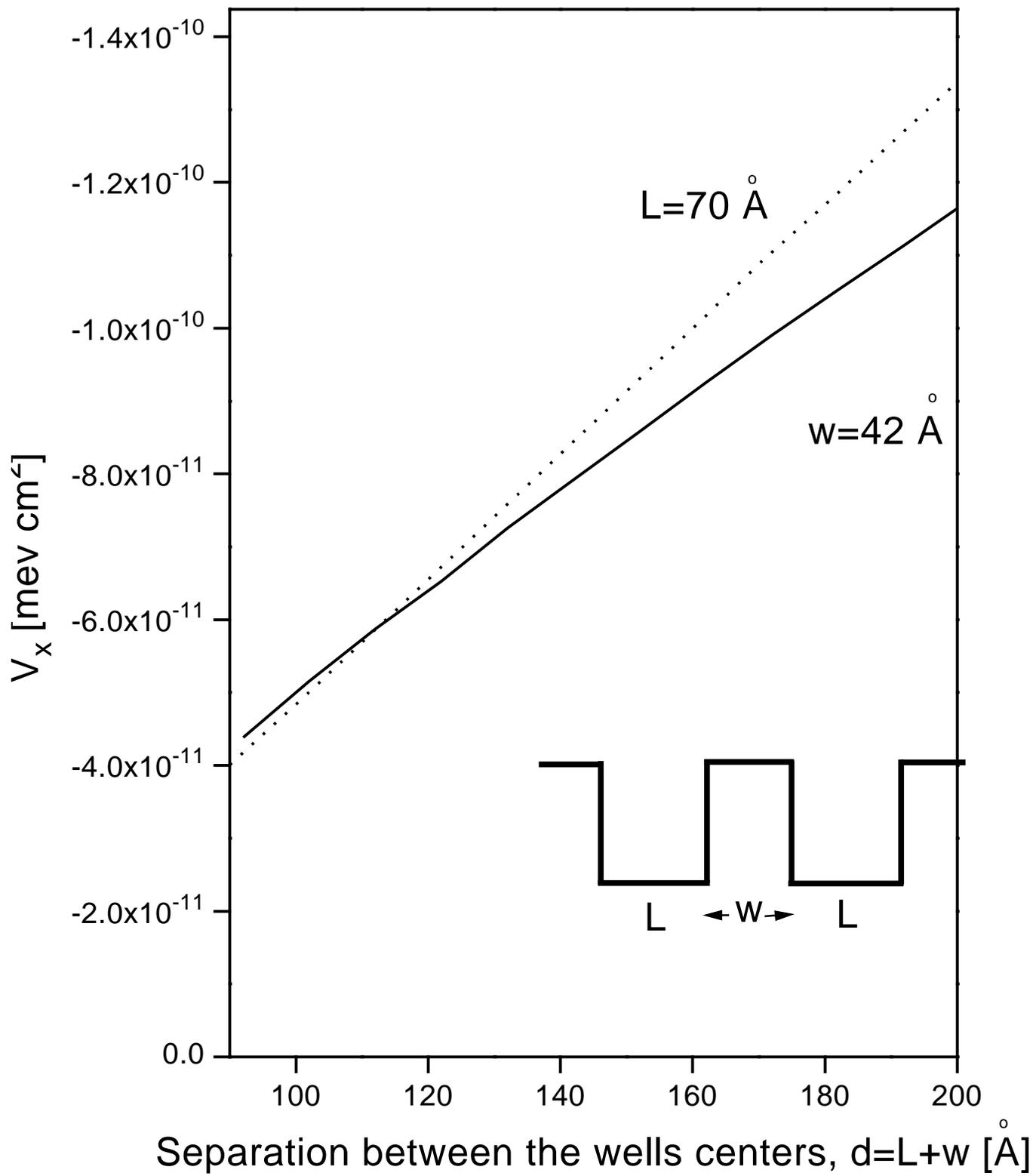
\begin{figure}
\caption{The dependence of the exchange coefficient,
$V_{x}=U^{(x)}-2|\epsilon_b|A$, on the separation between the wells,
$d=w+L$. The continuous line is for fixed barrier width, $w=42$ \AA, and
changing wells width. The dashed line is for fixed wells widths, $L=70$ \AA,
and changing the barrier width.}
\label{x}
\end{figure}

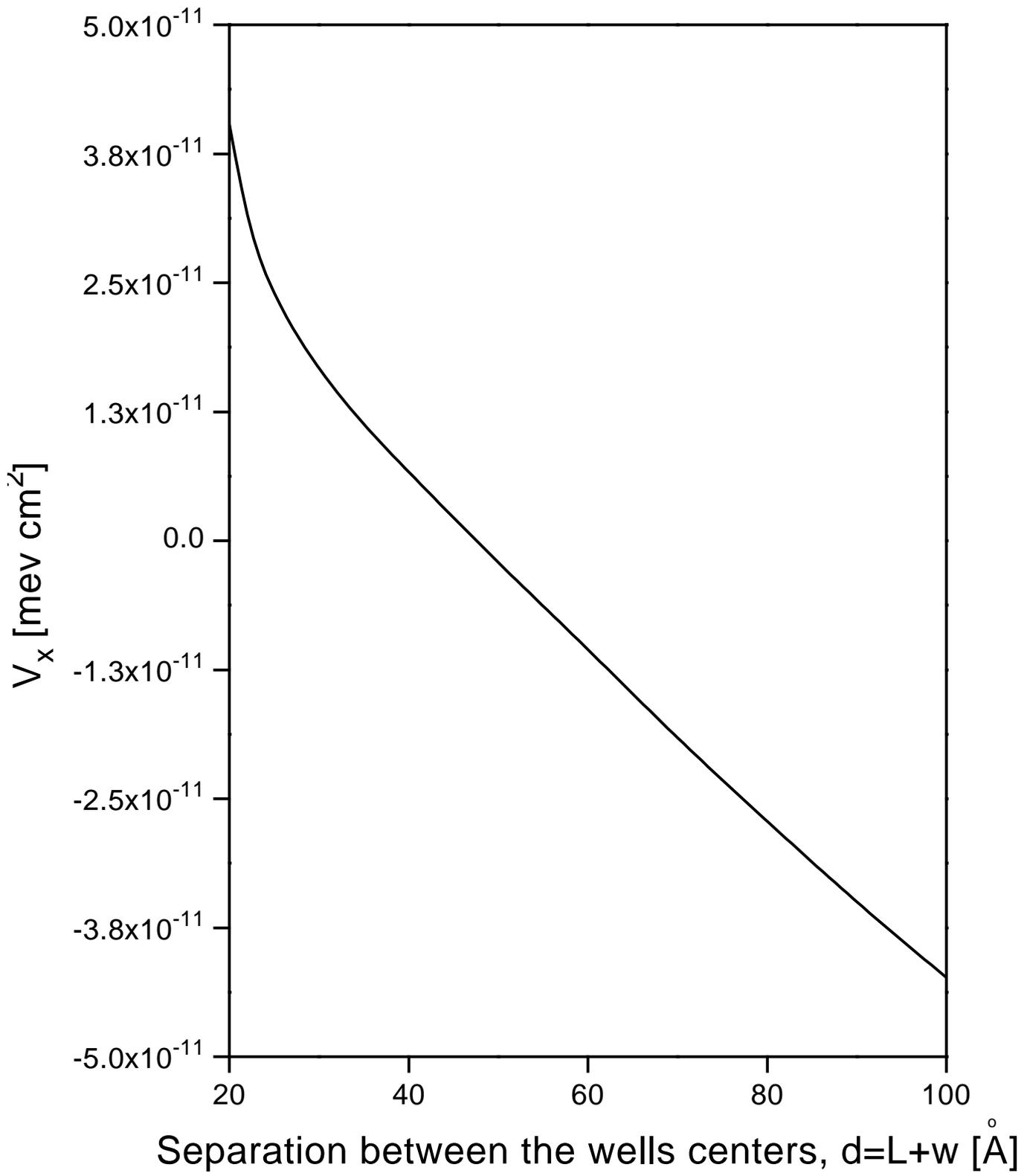
\begin{figure}
\caption{The dependence of the exchange coefficient, $V_{x}$, on the
separation between the wells, $d=w+L$, where $w=10$ \AA \ and $L$ is
changing. $V_{x}$ becomes positive for a small enough separation between the
wells.}
\label{x1}
\end{figure}

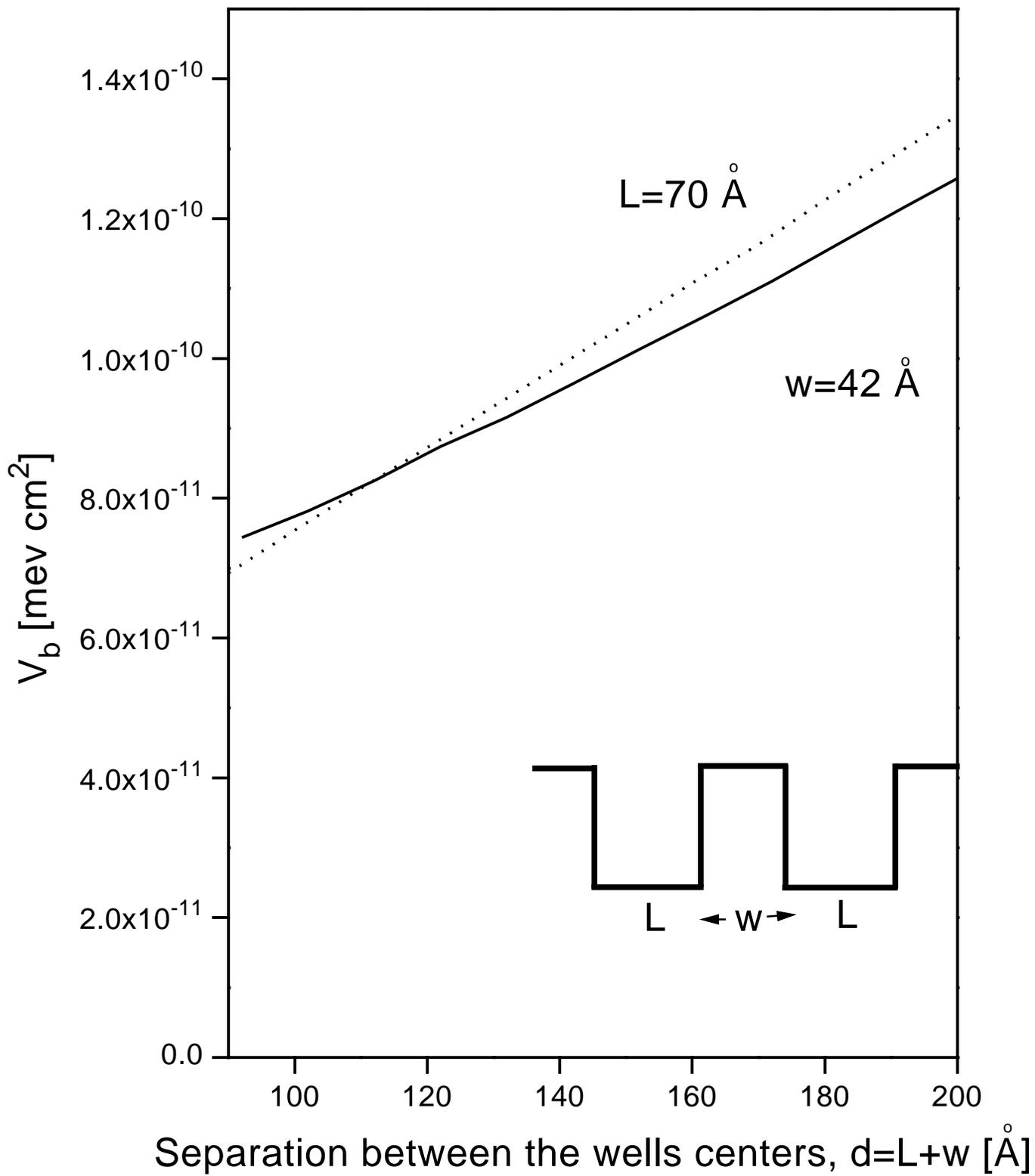
\begin{figure}
\caption{The dependence of the coefficient, $V_{b}$, on the
separation between the wells, $d=w+L$. The continuous line is for
fixed barrier width, $w=42$ \AA, and changing wells width. The
dashed line is for fixed wells widths, $L=70$ \AA, and changing
the barrier width.}
\label{bs}
\end{figure}

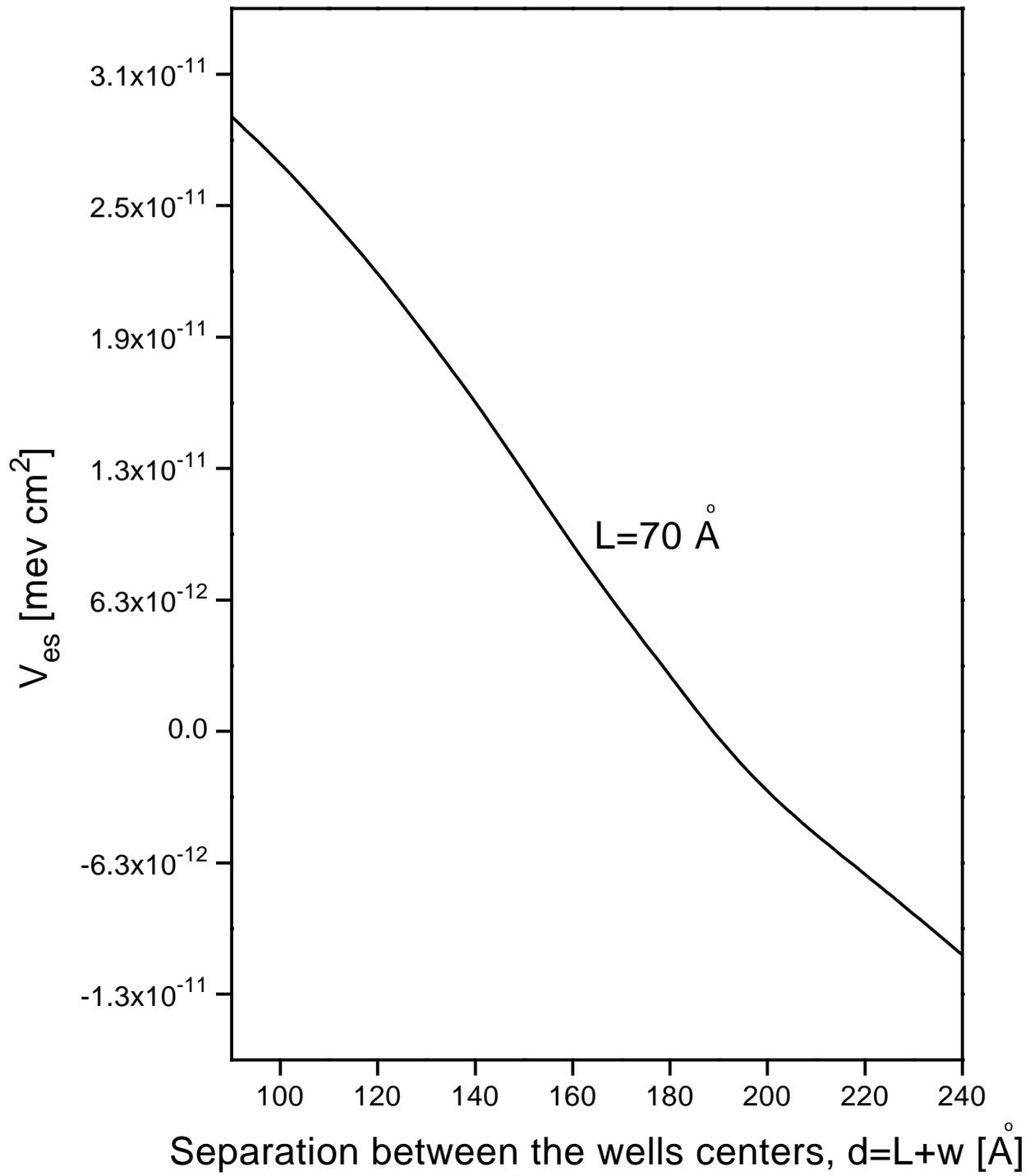
\begin{figure}
\caption{The dependence of the energy splitting parameter, $V_{es}$, on the
separation between the wells, $d=w+L$, where $L=70$ \AA \ and $w$ is
changing.} \label{es}
\end{figure}

We present numerically calculated $V_{x}$, $V_{b}$ and $V_{es}$ for a
symmetric Al$_x$Ga$_{1-x}$As / GaAs / Al$_x$Ga$_{1-x}$As / GaAs /
Al$_x$Ga$_{1-x}$As coupled quantum wells structure with the well widths
$L_{e}=L_{h}=L$. In the numerical calculations we use the electron effective
mass $m_e=0.067 m_0$, the hole effective masses $m_\perp=0.45 m_0$ and
$m_\parallel=0.126 m_0$ (here $m_{0}$ is the free electron mass), and the
dielectric constant $\kappa=12.5$. In the previous paper \cite{smadar} we
have shown that the exciton wave function depends mainly on the distance
between the centers of the wells and is not very sensitive to details of
their geometry. We can expect the same from the parameters $V_{x}$, $V_{b}$
and $V_{es}$. For this reason in Figs. \ref{x} - \ref{es} we present the
dependence of $V_{x}$, $V_{b}$ and $V_{es}$ on this distance, $d=w+L$. To
show the sensitivity of these parameters to the barrier width and the well
width separately we give on Figs. \ref{x} and \ref{bs} two curves, one for a
given $w$ (solid line) and the other for a given $L$ (dotted line).

The parameter $V_x$ can be positive or negative depending on the separation
between the wells $d$. To demonstrate this we presented its dependence on
$d$ for different ranges in Figs. \ref{x} and \ref{x1}. The reason for the
change of the sign is that $V_{x}$ contains two contributions of different
sign, the negative quantity $U^{(x)}$ and the positive quantity $-2 A
\epsilon_b$, Eq. (\ref{eq:deh.heg.5}) (we neglect $V_{2}$). At small
distances between the wells the second term dominates. When the separation
between the wells increases $U^{(x)}$ is just weakly affected while the
binding energy $\epsilon_{b}$ decreases, which leads to the change of the
sign.

In Fig. \ref{bs} we present the parameter $V_{b}$ which characterizing the
overall shift of the exciton luminescence line. It is the combination of the
direct interaction and the exchange term, $V_{x}$, Eq. (\ref{eq:mfa.2}). The
direct interaction dominates and $V_{b}$ is always positive, which leads to
the blue shift of the line.

To demonstrate the possibility of paramagnetic and ferromagnetic phases of
the exciton gas we present the dependence of $V_{es}$ on $d$ in Fig.
\ref{es}. We see that $V_{es}$ is a decreasing function of the separation. It
becomes negative at large separation, which corresponds to ferromagnetic
phase.

In a single infinite quantum well the electron and hole wave functions
describing the confinement in the well are equal, and according to Eq.
(\ref{eq:deh.2eh.17}) $U^{(d)}(q)$ is identically
zero.\cite{Tejedor96,Ciuti98}  For an estimate of the exchange matrix
element $V_{x}$ in a single-well we can use two-dimensional model for which
the exciton wave function is known (see, e.g.,
Ref.\onlinecite{Schmitt-Rink}) and which is typically used for such
estimates.\cite{Amand94,Schmitt-Rink,Tejedor96,Ciuti98} In this case the
wave function is a simple exponent, the absolute value of the binding energy
is $\epsilon_{b}=2\mu e^{4}/\hbar^{2}\kappa^{2}$ and
\begin{mathletters}
\begin{eqnarray}
\epsilon_{b} A & = & - {8\pi \over 5} \ {\hbar^{2} \over \mu} \ ,
\label{eq:mfa.10a} \\
V_{x} & = & {4\pi\hbar^{2} \over \mu}
    \left(1 - {315 \pi^{2} \over 4096}\right) \approx
    3.03 \ {\hbar^{2} \over \mu} \ .
\label{eq:mfa.10b}
\end{eqnarray}
\label{eq:mfa.10}
\end{mathletters}
The corresponding coefficient in the expression for the splitting,
(\ref{eq:mfa.6b}) appears to be twice larger than in
Ref.\onlinecite{Tejedor96}. This difference comes from a different numerical
factor in the Hamiltonian.

\section{Exciton-exciton relaxation time.}
\label{esr}

In this section we study exciton gas relaxation due to exciton-exciton
collisions. In these collisions, the excitons change their momenta and can
change their spins, however, the sum of exciton spins is conserved. For
further calculation it is convenient to introduce the notation
$M_{s_{1}s_{2}}^{s_{3}s_{4}}(q)$ for the scattering matrix element that
describes the scattering from the state with spins $s_{1}$ and $s_{2}$ to the
state with spins $s_{3}$ and $s_{4}$ ($s_{1}+s_{2}=s_{3}+s_{4})$ with the
transferred momentum $\hbar q$. In the Born approximation, according to
Hamiltonian (\ref{eq:deh.heg.6}) the spin dependence of the matrix element
is reduced to the separation of three cases. If the sum of exciton spins is
nonzero then the collision matrix element
\begin{mathletters}
\begin{equation}
M_{s_{1}s_{2}}^{s_{1}s_{2}}(q) = U^{(d)}(q) + V_{x} \ ,
    \hspace{1cm} s_{2} + s_{1} \neq 0 \ .
\label{eq:rt.1a}
\end{equation}
If the total spin is zero and spins in the final state are different from
spins in the initial state then
\begin{equation}
M_{s_{1}-s_{1}}^{s_{2}-s_{2}}(q) = {V_{x} \over 2} \ ,
    \hspace{1cm} s_{2} \neq s_{1} \ .
\label{eq:rt.1b}
\end{equation}
If the total spin is zero and initial and final spins are the same then
\begin{equation}
M_{s-s}^{s-s}(q) = U^{(d)}(q) - {V_{x} \over 2} \ .
\label{eq:rt.1c}
\end{equation}
\label{eq:rt.1}
\end{mathletters}

For 2D scattering, however, the Born approximation at low energies is not
satisfactory and the matrix element can be strongly
renormalized.\cite{Schick71,Lapidus82,Popov} For a spin independent
interaction between particles, $U(\vec{r})$, the renormalization is reduced
to the division of the Born matrix element by
\begin{equation}
1 - {M \over \hbar^{2}} \int
        \left[
    {1 \over 2i} + {1 \over \pi} \left(C + \ln{Kr \over 2}\right)
        \right]
    U(\vec{r}) d\vec{r} \ ,
\label{eq:rt.2}
\end{equation}
where $\hbar^2 K^2/M$ is the kinetic energy in the center of mass reference
frame and $C$ is the Euler constant. We don't calculate the renormalization
for a spin dependent potential since we use Eq. (\ref{eq:rt.2}) only for
estimates.

The spin relaxation in the exciton gas is usually described by simplified
kinetic equations that ignore exciton momentum
distribution.\cite{Amand97,Maialle00}. The Bolzmann equation that describes
both spin and momentum relaxation has the form
\begin{eqnarray}
{\partial n_{\vec{K}s}\over\partial t} & = &
    {2\pi \over \hbar} \sum_{s_{1}\neq-s}
    \int \left|M_{ss_{1}}^{ss_{1}}(q)\right|^{2}
        \Big[
    (n_{\vec{K}s} + 1)(n_{\vec{K}_{1}s_{1}} + 1)
    n_{\vec{K}+\vec{q}s}n_{\vec{K}_{1}-\vec{q}s_{1}}
\nonumber \\ && \hspace{2.8cm} -
    (n_{\vec{K}+\vec{q}s} + 1)(n_{\vec{K}_{1}-\vec{q}s_{1}} + 1)
    n_{\vec{K}s}n_{\vec{K}_{1}s_{1}}
        \Big]
\nonumber \\ && \hspace{2.8cm} \times
    \delta  \left(
    E_{K} + E_{K_{1}} - E_{\vec{K}+\vec{q}} - E_{\vec{K}_{1}-\vec{q}}
            \right)
    {d^{2}K_{1} \over (2\pi)^{2}} \ {d^{2}q \over (2\pi)^{2}}
\nonumber \\ & + &
    {2\pi \over \hbar} \sum_{s_{1}}
    \int \left|M_{s-s}^{s_{1}-s_{1}}(q)\right|^{2}
        \Big[
    (n_{\vec{K}s} + 1)(n_{\vec{K}_{1}-s} + 1)
    n_{\vec{K}+\vec{q}s_{1}}n_{\vec{K}_{1}-\vec{q}-s_{1}}
\nonumber \\ && \hspace{2.8cm} -
    (n_{\vec{K}+\vec{q}s_{1}} + 1)(n_{\vec{K}_{1}-\vec{q}-s_{1}} + 1)
    n_{\vec{K}s}n_{\vec{K}_{1}-s}
        \Big]
\nonumber \\ && \hspace{2.8cm} \times
    \delta  \left(
    E_{K} + E_{K_{1}} - E_{\vec{K}+\vec{q}} - E_{\vec{K}_{1}-\vec{q}}
            \right)
    {d^{2}K_{1} \over (2\pi)^{2}} \ {d^{2}q \over (2\pi)^{2}} \ .
\label{eq:rt.3}
\end{eqnarray}

Numerical solution of Eq. (\ref{eq:rt.3}) is much more difficult than its
simplified versions that ignore momentum distribution. In general, the
relaxation according to Eq. (\ref{eq:rt.3}) cannot be exactly described by a
relaxation time. To characterize the relaxation rate it is possible,
nevertheless, to introduce an inverse relaxation time as the coefficient for
$-n_{\vec{K}s}$ in the collision operator
\begin{eqnarray}
{1 \over \tau_{K s}} & = &
    {2\pi \over \hbar} \sum_{s_{1}\neq-s}
    \int \left|M_{ss_{1}}^{ss_{1}}(q)\right|^{2}
        \Big[
    (n_{\vec{K}+\vec{q}s} + n_{\vec{K}_{1}-\vec{q}s_{1}} + 1)
    n_{\vec{K}_{1}s_{1}} -
    n_{\vec{K}+\vec{q}s}n_{\vec{K}_{1}-\vec{q}s_{1}}
        \Big]
\nonumber \\ && \hspace{2.8cm} \times
    \delta  \left(
    E_{K} + E_{K_{1}} - E_{\vec{K}+\vec{q}} - E_{\vec{K}_{1}-\vec{q}}
            \right)
    {d^2{K}_{1} \over (2\pi)^{2}} \ {d^2{q} \over (2\pi)^{2}}
\nonumber \\ & + &
    {2\pi \over \hbar} \sum_{s_{1}}
    \int \left|M_{s-s}^{s_{1}-s_{1}}(q)\right|^{2}
        \Big[
    (n_{\vec{K}+\vec{q}s_{1}} + n_{\vec{K}_{1}-\vec{q}-s_{1}} + 1)
    n_{\vec{K}_{1}-s} -
    n_{\vec{K}+\vec{q}s_{1}}n_{\vec{K}_{1}-\vec{q}-s_{1}}
        \Big]
\nonumber \\ && \hspace{2.8cm} \times
    \delta  \left(
    E_{K} + E_{K_{1}} - E_{\vec{K}+\vec{q}} - E_{\vec{K}_{1}-\vec{q}}
            \right)
    {d^2{K}_{1} \over (2\pi)^{2}} \ {d^2{q} \over (2\pi)^{2}}
\label{eq:rt.4}
\end{eqnarray}

This expression contains both linear and quadratic terms in exciton
occupation numbers. Respectively, an order of magnitude estimate of the
relaxation time contains linear and quadratic terms in the exciton
concentration $n$,
\begin{equation}
{1 \over \tau_{Ks}} \sim
    {n^2 + n K^2 \over \hbar (\hbar^{2}K^{2}/\mu)} \
    \left|M_{ss_{1}}^{ss_{1}}(K)\right|^{2} \ .
\label{eq:rt.5}
\end{equation}

The relaxation time $\tau_{Ks}$ characterizes energy and momentum relaxation
in the exciton gas. The same time characterizes also a partial spin
relaxation. The spin relaxation due to collisions cannot be complete because
of the total spin conservation in collisions. Formally it is described by
the identity
\begin{equation}
{\partial\over\partial t} \ (n_{s} - n_{-s}) = 0 \ ,
\label{eq:rt.6}
\end{equation}
that follows from Eq. (\ref{eq:rt.3}). Here $n_{s}$ is the concentration of
excitons with spin $s$. That is, the only spin relaxation due to collisions
is the relaxation between dark excitons, (excitons with spin $\pm2$), and
bright excitons, (excitons with spin $\pm1$) (see also
Ref.\onlinecite{Ciuti98}). It immediately follows from here that spin
relaxation in the exciton gas is characterized by few relaxation times which
correspond to different relaxation mechanisms. If the exciton concentration
is not very small then the fastest relaxation is the dark - bright exciton
relaxation characterized by $\tau_{Ks}$. Complete relaxation can take place
due to processes that involve D'yakonov - Perel mechanism of electron spin
relaxation, light - heavy hole mixing, or electron - hole
exchange.\cite{sham} All these mechanisms contain their respective small
coupling constants and are activated by scattering. So, if the main exciton
scattering mechanism is exciton-exciton scattering then the complete spin
relaxation has to be much slower than dark - bright exciton relaxation. The
relation between the relaxation times can be different for a very small
exciton concentration when other scattering mechanisms, e.g., phonon,
impurity or surface roughness scattering are important.

\section{Discussion}
\label{sec:dis}

In this section we compare our results with a few experiments. We consider
the shift of the exciton luminescence line, the energy splitting between
exciton with different spins, the polarization of the exciton gas, and the
exciton-exciton scattering time.

As we already mentioned in the introduction, the density dependence of the
exciton luminescence line shift,
\cite{Hulin1,Hulin2,Dareys93,Amand94,Butov,Snoke} the time dependence of
line spin splitting and the luminescence depolarization
\cite{Damen91,Vina96,Jeune98,Vina99,Baylac95,Amand97} proved that all these
phenomena come from exciton-exciton interaction. A theoretical study of
these phenomena has been done by Fern\'{a}ndez-Rossier {\it et
al}.\cite{Tejedor96,Tejedor97}, Ciuti {\it et al}. \cite{Ciuti98}, Amand
{\it et al.}\cite{Amand97}, and Maialle {\it et al.}\cite{Maialle00}. Here
we consider only some features of the experiments which have not found a
clear explanation so far.

The density dependent blue shift of the exciton line in
GaAs/Al$_{x}$Ga$_{1-x}$As symmetric coupled quantum wells, where electrons
and holes are spatially separated by external gate voltage, has been
recently measured by Butov {\it et al.}\cite{Butov} and Negoita {\it et al}.
\cite{Snoke}. Butov {\it et al}. detected blue shift of 1.6 meV at zero
magnetic field for wells width $L=80$ \AA \ and barrier width of $w=40$ \AA.
They used the plate capacitor expression for the direct interaction,
neglecting the exchange, to calculate the exciton density of $n=9\times
10^{9}$ cm$^{-2}$. To compare our results with their measurements we
calculate the concentration according to Eq. (\ref{eq:mfa.3}) and check the
importance of the corrections. For the described geometry we obtain
$V_{b}=8.7 \times 10^{-11}$ meV$\cdot$ cm$^{-2}$. The temperature of this
experiment, $50$ mK, is so low that we can use Eq. (\ref{eq:mfa.6}) instead
of Eq. (\ref{eq:mfa.3}), where $U^{(d)}(0)=1.5\times 10^{-10}$ meV$\cdot$
cm$^{-2}$. Assuming equal concentration of excitons with different spins,
$n_{s}=n_{-s}=n/4$, we have $\delta \epsilon_s= V_{b} n + U^{(d)}(0) n/4$
and for $\delta \epsilon_s=1.6$ meV we obtain $n=1.3\times 10^{10}$
cm$^{-2}$, which is close to the concentration obtained from the simple
plate capacitor expression without the exchange correction.

The comparison with the results of Negoita {\it et al}.\cite{Snoke} is more
interesting because there the excitation concentration was measured from the
excitation intensity, independently of the blue shift. The measurements were
made in symmetric coupled quantum wells, where the wells width was $L=60$
\AA \ and the barrier width was $w=42$ \AA. The lattice temperature was 2 K
and the concentration was in the range $10^9-10^{12}$ cm$^{-2}$. For low
density, linear blue shift of $5\times 10^{-11}$ meV $\cdot$ cm$^{2}$, was
observed. For this geometry we have $V_{b}=7.8\times 10^{-11}$ meV $\cdot$
cm$^{2}$. In the case, $n_{s}=n_{-s}=n/4$, with
$U^{(d)}(0)=1.3\times10^{-10}$ meV$\cdot$ cm$^{-2}$ we get from Eq.
(\ref{eq:mfa.6}) linear shift of $1.1\times10^{-10}$ meV $\cdot$ cm$^{2}$.
The difference between this value and the experimental one can result from
our assumption that electrons and holes are completely confined in separate
wells which increases the direct interaction $U^{(d)}$. Another reason can be
the presence of free carriers in the experiment, which screen the Coulomb
potential and make exciton-exciton interaction weaker. The later possibility
is supported by the luminescence line width which is larger than the exciton
binding energy.

In the same structure, in a weak magnetic field, Snoke {\it et
al}.\cite{Snoke1} observed a red shift of the exciton luminescence line
which grew with the gate voltage that separated electrons and holes. The
most striking result is that the red shift reaches values of 10 or 20 meV
(depending on the gate voltage) at magnetic field around 1 T. Such a
magnetic field is not strong enough to induce a significant blue diamagnetic
line shift.\cite{Zhao,Oettinger} The explanation that we suggest is based on
a very narrow line of the pumping laser. Even a weak magnetic field can
split the exciton lines with different polarizations so much that they go
away from the resonance with the pumping laser. This leads to a reduction of
the exciton density, resulting, according to Eq. (\ref{eq:mfa.6}), with a red
shift of the luminescence line. An increase of the external electric field
increases the separation between electrons and holes leading to larger
values of the coefficients $V_{b}$ and $U^{d}(0)$. As a result, the red shift
also increases, as it is observed in the experiment. The effect is symmetric
to the direction of the magnetic field, which is also in agreement with the
experiment. The magnitude of the effect depends only on the absolute
concentration change, $\Delta n$, so the relative change $\Delta n/n$ in the
luminescence intensity, can be small.

Another phenomenon related to exciton-exciton interaction is a spontaneous
energy splitting between excitons with opposite spins. A typical
experimental way to produce a polarized exciton gas, (i.e., a gas where
$n_{-s}\neq n_{s}$) is pumping by polarized light. In
Refs.\onlinecite{Damen91,Vina96,Jeune98,Vina99}, the energy spin splitting
was measured in multiple quantum wells where the electrons and holes are in
the same well. This corresponds to zero separation between the carriers and
therefore zero $U^{(d)}(0)$ and positive $V_{es}$. In all the experiments the
spin majority excitons had higher energy than the minority and the
difference increased with the density, as we would expect from positive
$V_{es}$. Another evidence for $V_{es}$ being positive in multiple quantum
well systems is the depolarization of the initially polarized exciton gas,
that was reported by different authors
\cite{Dareys93,Damen91,Jeune98,Vina99,Amand97}. According to the results of
Sec. \ref{sec:mfa}, when $V_{es}$ is positive the system is paramagnetic,
and the minimal energy of the system is reached when the exciton gas is
depolarized.

In a double well structure Aichmayr {\it et al}.\cite{Mendez} detected an
energy spin splitting dependence on the gate voltage that separated
electrons and holes. As the voltage increased, the energy splitting
decreased from 4 meV to zero. This behavior corresponds to Eq.
(\ref{eq:mfa.6b}), where with the increase of the electron and the hole
separation, the coefficient $V_{es}$, being positive, decreases to zero (see
Fig. \ref{es}). A more detailed consideration, with the help of Eq.
(\ref{eq:mfa.6a}), can describe a different behaviour of the minority and
majority exciton luminescence lines which is presented in Fig.2 of
Ref.\onlinecite{Mendez}. At low gate voltage the lines are split nearly
symmetrically (the shift of the majority line is positive while the shift of
the minority line is negative) with respect to the value to which both of
them relax with the time constant $\tau_{sd}=180$ ps. With increase of the
voltage the shift of the majority line does not change while the shift of
the minority line decreases, becomes positive and at a very high voltage the
splitting disappears. First of all it is necessary to note that the
luminescence decay time (400 ps for the low voltage and 1000 ps for the high
voltage) is a few times larger than $\tau_{sd}$. The luminescence decay
characterizes the decrease of the exciton concentration and the
comparatively small value of $\tau_{sd}$ means that the reason the splitting
relaxation is not decrease of the exciton concentration but spin relaxation,
probably due to light and heavy hole mixing. So the line shift due to
exciton polarization under the condition of constant total concentration can
be calculated according to
$\Delta\epsilon_{s}=V_{x}(n_{s}-n_{-s})+U^{(d)}(0)n_s$. At low gate voltage
the separation between electrons and holes is small, the direct interaction
$U^{(d)}(0)$ is negligible and the shifts of the majority (+1) and minority
(-1) lines are $\Delta\epsilon_{\pm1}=\pm V_{x}(n_{+1}-n_{-1})$. This is a
symmetric shift in agreement with the experiment. With increase of the
voltage the separation between electrons and holes grows leading to growth
of $U^{(d)}(0)$ and decrease of $V_{x}$. Given the concentrations, at some
intermediate voltage $V_{x}=U^{(d)}(0)n_{-1}/(n_{+1}-n_{-1})$ and then
$\Delta\epsilon_{+1}=U^{(d)}(0)(n_{+1}+n_{-1})$ while
$\Delta\epsilon_{-1}=0$. This corresponds to Fig.2b of
Ref.\onlinecite{Mendez}. At high voltage $V_{x}$ is negative, and if
$V_{x}\approx-U^{(d)}(0)/2$ then
$\Delta\epsilon_{\pm1}=U^{(d)}(0)(n_{+1}+n_{-1})/2$ which corresponds to
Fig.2c of Ref.\onlinecite{Mendez}. That is Eq.(\ref{eq:mfa.6a}) completely
describes the behavior of both minority and majority lines. It makes sense to
note that $U^{(d)}(0)$ at the intermediate field is smaller than at high
field so the majority line does not shift much, which also corresponds to
the experiment.

The last physical phenomena we want to discuss here is the exciton-exciton
relaxation time. For the estimate of the relaxation time we make use of Eq.
(\ref{eq:rt.5}). The time necessary for an exciton to emit or absorb a
phonon is around hundreds of ps \cite{Takagahara} which is much longer than
short luminescence relaxation times with electrons and hole in the same well
(a few tens of ps). That means that under the condition of resonant
excitation the exciton momentum is around that of an exciting photon,
$K\sim2\times10^{5}$ cm$^{-1}$. So for concentrations of the order or
smaller than $4\times10^{10}$ cm$^{-2}$ we can use for an estimate only the
second term in the numerator of Eq. (\ref{eq:rt.5}). Since the relevant
experiments have been done in multi quantum wells structures, where
electrons and holes are confined in the same layer, we make use of the 2D
model that we have discussed in the end of Sec. \ref{sec:mfa} and estimate
the interaction matrix element according to Eq. (\ref{eq:mfa.10b}). In the
renormalization factor (\ref{eq:rt.2}) only the logarithmic term can be
important and it is $\sim1-(M/\pi\hbar^{2})V_{x}\ln(Ka)\approx8.7$ for the
exciton radius $a\approx80$ \AA. This leads to $1/\tau\approx0.12(\hbar
n/\mu)$ and $\tau\sim8$ ps which is close to the experimental spin
relaxation time measured by Le Jeune {\it et al}.\cite{Jeune98}, Baylac {\it
et al}.
\cite{Baylac95}, and Amand {\it et al}.\cite{Amand97}.

Wang {\it et al}. \cite{Wang} measured the exciton momentum
relaxation rate in the concentration region  between $10^9$ to
$1.5 \times 10^{10}$ cm$^{-2}$ in multiple quantum wells of 130
\AA~width. They obtained the relaxation rate that grew with the
concentration from 0.5 to 2 ps$^{-1}$  that is about an order of magnitude
larger than Eq. (\ref{eq:rt.5}) gives. Such a big difference cannot be
attributed to a deviation for 2D model for exciton. It is likely that a
contribution of other elastic scattering mechanisms (e.g., surface
roughness) was substantial in this experiment.

\section{Conclusion}
\label{con}
We derived the Hamiltonian of exciton gas in quantum wells by the projection
of the electron-hole plasma Hamiltonian to exciton states and expansion in a
small exciton density. Matrix elements of the exciton Hamiltonian are
expressed in terms of a single exciton wave function which is not modified by
exciton-exciton interaction and is rather sensitive to the geometry of the
heterostructure. With the help of the exciton Hamiltonian we estimated the
blue shift and spin splitting of the exciton luminescence line and their
dependence on the heterostructure parameters. We also wrote down the
Boltzmann equation for excitons and estimated the energy and spin relaxation
time resulting from the exciton-exciton scattering. We succeeded to give an
explanation to some recent experimental results that have not been explained
so far.

\section{Acknowledgements}

The research was supported by The Israel Science Foundation founded by the
Israel Academy of Sciences and Humanities.

\appendix
\section{Calculation of matrix elements}
\label{cme}

In this Appendix we present the detailed calculation of the
matrix elements that determine the exciton-exciton pair
interaction. The Appendix contains four subsections where the
overlap integral ${\cal
A}_{\vec{K}_{1}s_{1},\vec{K}_{2}s_{2};\vec{K}_{3}s_{3},\vec{K}_{4}s_{4}}$,
the direct part of the matrix element
$H_{\vec{K}_{1}s_{1},\vec{K}_{2}s_{2};\vec{K}_{3}s_{3},\vec{K}_{4}s_{4}}^{(d)}$,
(\ref{eq:deh.2eh.10}), the exchange part of the matrix element
$H_{\vec{K}_{1}s_{1},\vec{K}_{2}s_{2};\vec{K}_{3}s_{3},\vec{K}_{4}s_{4}}^{(x)}$,
(\ref{eq:deh.2eh.11}), and the contribution from the excited
states, $H_{\vec{K}_{1}s_{1},\vec{K}_{2}s_{2},
        \vec{K}_{3}s_{3},\vec{K}_{4}s_{4}}^{(exex)}$,
are calculated. In this calculation we omit $z$-dependent part of
the kinetic and potential energy in ${\cal H}_{2eh}$. This part
gives only the size quantization energy of free particles which
is our energy reference point.

\subsection{Overlap integral}
\label{sec:cme.oi}

The substitution of functions (\ref{eq:deh.2eh.4}) in the definition
(\ref{eq:deh.rh.17}) gives
\begin{eqnarray}
&& {\cal A}_{\vec{K}_{1}s_{1},\vec{K}_{2}s_{2};
        \vec{K}_{3}s_{3},\vec{K}_{4}s_{4}} =
    {1 \over S} \ \delta_{\vec{K}_{1}+\vec{K}_{2},\vec{K}_{3}+\vec{K}_{4}}
\nonumber \\ && \hspace{1cm} \times
        \left(
    \delta_{s_{1e}s_{4e}}\delta_{s_{2e}s_{3e}}
    \delta_{s_{1h}s_{3h}}\delta_{s_{2h}s_{4h}}
    A_{\vec{K}_{1}\vec{K}_{2},\vec{K}_{3},\vec{K}_{4}} +
    \delta_{s_{1e}s_{3e}}\delta_{s_{2e}s_{4e}}
    \delta_{s_{1h}s_{4h}}\delta_{s_{2h}s_{3h}}
    A_{\vec{K}_{1}\vec{K}_{2},\vec{K}_{4},\vec{K}_{3}}
        \right) ,
\label{eq:cme.oi.1}
\end{eqnarray}
$s_{je}$ ($s_{jh}$) is the electron (hole) spin of the exciton
with spin $s_{j}$, and
\begin{eqnarray}
&&  A_{\vec{K}_{1}\vec{K}_{2},\vec{K}_{3},\vec{K}_{4}} =
    - \int
    d\vec{R} d\vec{r}_{ee} d\vec{r}_{hh} \ \exp
        \left[
    - i(\vec{K}_{1} - \vec{K}_{4}) \vec{r}_{ee}
    - i(\vec{K}_{1} - \vec{K}_{3}) \vec{r}_{hh}
        \right]
\nonumber \\
&& \times
    \phi\left(\vec{R} + {\vec{r}_{ee} - \vec{r}_{hh} \over 2}\right)
    \phi\left(\vec{R} - {\vec{r}_{ee} - \vec{r}_{hh} \over 2}\right)
    \phi\left(\vec{R} - {\vec{r}_{ee} + \vec{r}_{hh} \over 2}\right)
    \phi\left(\vec{R} + {\vec{r}_{ee} + \vec{r}_{hh} \over 2}\right) \ .
\label{eq:cme.oi.2}
\end{eqnarray}
The overlap integral (\ref{eq:cme.oi.2}) has an obvious property,
$A_{\vec{K}_{1}\vec{K}_{2},\vec{K}_{3},\vec{K}_{4}}=
    A_{\vec{K}_{1}\vec{K}_{2},\vec{K}_{4},\vec{K}_{3}}$.
The reduction of the overlap integral to this form has been made with the
help of the relative electron-hole coordinates
\begin{mathletters}
\begin{eqnarray}
    \vec{r}_{ee} & = & \vec{r}_{e1} - \vec{r}_{e2} \ ,
    \hspace{1cm} \vec{r}_{hh} =  \vec{r}_{h1} - \vec{r}_{h2} \ ,
\label{eq:cme.oi.3a} \\
    \vec{R} & = &
    (\vec{r}_{e1}+\vec{r}_{e2}-\vec{r}_{h1}-\vec{r}_{h2})/2 \ .
\label{eq:cme.oi.3b}
\end{eqnarray}
\label{eq:cme.oi.3}
\end{mathletters}
The characteristic values of $R$, $r_{ee}$, and $r_{hh}$ in integral
(\ref{eq:cme.oi.2}) are of the order of the exciton radius $a$. On the other
hand, inequality (\ref{eq:deh.2}) is equivalent to $aK\ll1$. This means that
all exponential factors in the integrand can be replaced by unity and
\begin{eqnarray}
&&  A_{\vec{K}_{1}\vec{K}_{2},\vec{K}_{3},\vec{K}_{4}} = A \equiv
    - \int
    d\vec{R} d\vec{r}_{ee} d\vec{r}_{hh} \
\nonumber \\
&& \times
    \phi\left(\vec{R} + {\vec{r}_{ee} - \vec{r}_{hh} \over 2}\right)
    \phi\left(\vec{R} - {\vec{r}_{ee} - \vec{r}_{hh} \over 2}\right)
    \phi\left(\vec{R} - {\vec{r}_{ee} + \vec{r}_{hh} \over 2}\right)
    \phi\left(\vec{R} + {\vec{r}_{ee} + \vec{r}_{hh} \over 2}\right) \ .
\label{eq:cme.oi.4}
\end{eqnarray}
The substitution here of the Fourier transform of the wave
function, (\ref{eq:deh.2eh.8}), immediately leads to Eq.
(\ref{eq:deh.2eh.7}).

\subsection{The direct part.}
\label{sec:cme.dir}

After the summation over spin variables, Eq.
(\ref{eq:deh.2eh.10}) becomes
\begin{eqnarray}
H_{\vec{K}_{1}s_{1},\vec{K}_{2}s_{2},
        \vec{K}_{3}s_{3},\vec{K}_{4}s_{4}}^{(d)} & = &
    \delta_{s_{1},s_{3}}\delta_{s_{2},s_{4}}
    \int
    \psi_{\vec{K}_{1}}^{\ast}(\vec{r}_{e1}, \vec{r}_{h1})
    \psi_{\vec{K}_{2}}^{\ast}(\vec{r}_{e2}, \vec{r}_{h2})
    {\cal H}_{2eh}
\nonumber \\ && \hspace{0.5cm} \times
    \psi_{\vec{K}_{3}}(\vec{r}_{e1}, \vec{r}_{h1})
    \psi_{\vec{K}_{4}}(\vec{r}_{e2}, \vec{r}_{h2})
    d\vec{r}_{e1} d\vec{r}_{h1} d\vec{r}_{e2} d\vec{r}_{h2}
\nonumber \\ & + &
    \delta_{s_{1},s_{4}}\delta_{s_{2},s_{3}}
    \int
    \psi_{\vec{K}_{1}}^{\ast}(\vec{r}_{e1}, \vec{r}_{h1})
    \psi_{\vec{K}_{2}}^{\ast}(\vec{r}_{e2}, \vec{r}_{h2})
    {\cal H}_{2eh}
\nonumber \\ && \hspace{0.5cm} \times
    \psi_{\vec{K}_{4}}(\vec{r}_{e1}, \vec{r}_{h1})
    \psi_{\vec{K}_{3}}(\vec{r}_{e2}, \vec{r}_{h2})
    d\vec{r}_{e1} d\vec{r}_{h1} d\vec{r}_{e2} d\vec{r}_{h2} \ .
\label{eq:cme.dir.1}
\end{eqnarray}

The following part of the Hamiltonian (\ref{eq:deh.2eh.1})
\begin{eqnarray}
    - {{\hbar^2 \nabla_{e1}^2}\over{2 m_e}}
        - {{\hbar^2 \nabla_{h1}^2}\over{2 m_\parallel}}
        +u_{eh}(|\vec{r}_{e1}-\vec{r}_{h1}|)
    - {{\hbar^2 \nabla_{e2}^2}\over{2 m_e}}
        - {{\hbar^2 \nabla_{h2}^2}\over{2 m_\parallel}}
    +u_{eh}(|\vec{r}_{e2}-\vec{r}_{h2}|)
\label{eq:cme.dir.2}
\end{eqnarray}
is the sum of the Hamiltonians of two free excitons consisting of the pairs (e1,h1)
and (e2,h2). This part of the matrix element gives the sum of the two free excitons
energies, $E_{K_3}+E_{K_4}$ multiplied by the unit matrix, (\ref{eq:deh.2eh.5}).
Here $E_{K}=-\epsilon_b +\hbar^2 K^2/2 M$.
For the calculation of the other terms of the Hamiltonian, it is
convenient to change the in-plane variables of the integration to the center
of mass coordinate,
\begin{mathletters}
\begin{equation}
\vec{R}_{c} = [m_{e}(\vec{r}_{e1} + \vec{r}_{e2}) +
    m_{\parallel}(\vec{r}_{h1} + \vec{r}_{h2})] / 2M \ ,
\label{eq:cme.dir.3a}
\end{equation}
the distance between the exciton centers of mass,
\begin{equation}
\vec{\rho} = [m_{e}(\vec{r}_{e1} - \vec{r}_{e2}) +
    m_{\parallel}(\vec{r}_{h1} - \vec{r}_{h2})] / M \ ,
\label{eq:cme.dir.3b}
\end{equation}
and relative coordinates,
\begin{eqnarray}
&& \vec{r}_{1} = \vec{r}_{e1} - \vec{r}_{h1} \ , \hspace{0.5cm}
    \vec{r}_{2} = \vec{r}_{e2} - \vec{r}_{h2} \ .
\label{eq:cme.dir.3c}
\end{eqnarray}
\label{eq:cme.dir.3}
\end{mathletters}
Then the integration with respect to $\vec{R}_{c}$ results in Eq.
(\ref{eq:deh.2eh.12}), where
\begin{eqnarray}
&& U_{\vec{K}_{1}s_{1},\vec{K}_{2}s_{2},
        \vec{K}_{3}s_{3},\vec{K}_{4}s_{4}}^{(d)} =
    \delta_{s_{1}s_{3}} \delta_{s_{2}s_{4}} \int
    \exp    \left[
    i(- \vec{K}_{1} + \vec{K}_{2} + \vec{K}_{3} - \vec{K}_{4})
    {\vec{\rho} \over 2}
            \right]
\label{eq:cme.dir.4} \\
&& \times
        \left[
    u_{ee}(|\vec{r}_{e1} - \vec{r}_{e2}|) +
    u_{hh}(|\vec{r}_{h1} - \vec{r}_{h2}|) +
    u_{eh}(|\vec{r}_{e1} - \vec{r}_{h2}|) +
    u_{eh}(|\vec{r}_{e2} - \vec{r}_{h1}|)
        \right]
\nonumber \\ && \times
    \phi^{2}({r}_{1}) \phi^{2}({r}_{2})
    d\vec{r}_{1} d\vec{r}_{2} d\vec{\rho}
\nonumber \\ & + &
    \delta_{s_{1}s_{4}} \delta_{s_{2}s_{3}} \int
    \exp    \left[
    i(- \vec{K}_{1} + \vec{K}_{2} - \vec{K}_{3} + \vec{K}_{4})
    {\vec{\rho} \over 2}
            \right]
\nonumber \\ && \times
                \left[
    u_{ee}(|\vec{r}_{e1} - \vec{r}_{e2}|) +
    u_{hh}(|\vec{r}_{h1} - \vec{r}_{h2}|) +
    u_{eh}(|\vec{r}_{e1} - \vec{r}_{h2}|) +
    u_{eh}(|\vec{r}_{e2} - \vec{r}_{h1}|)
                \right]
\nonumber \\ && \times
    \phi^{2}({r}_{1})
    \phi^{2}({r}_{2})
    d\vec{r}_{1} d\vec{r}_{2} d\vec{\rho} \ .
\nonumber
\end{eqnarray}
With the help of the momentum conservation that is expressed by
$\delta_{\vec{K}_{1}+\vec{K}_{2},\vec{K}_{3}+\vec{K}_{4}}$ in
Eq. (\ref{eq:deh.2eh.12}), we come up with Eq. (\ref{eq:deh.2eh.13}) where
\begin{eqnarray}
U^{(d)}(q) & = & \int
    e^{-i\vec{q}\vec{\rho}}
        \Bigg[
    u_{ee}  \left(\left|
    \vec{\rho} + m_{\parallel} \ {\vec{r}_{1} - \vec{r}_{2} \over M}
            \right|\right) +
    u_{hh}  \left(\left|
    \vec{\rho} - m_{e} \ {\vec{r}_{1} - \vec{r}_{2} \over M}
            \right|\right)
\nonumber \\ && \hspace{1cm} +
    u_{eh}  \left(\left|
    \vec{\rho} + {m_{\parallel}\vec{r}_{1} + m_{e}\vec{r}_{2} \over M}
            \right|\right) +
    u_{eh}  \left(\left|
    \vec{\rho} - {m_{e}\vec{r}_{1} + m_{\parallel}\vec{r}_{2} \over M}
            \right|\right)
        \Bigg]
\nonumber \\ && \hspace{1cm} \times
    \phi^{2}({r}_{1}) \phi^{2}({r}_{2})
    d\vec{r}_{1} d\vec{r}_{2} d\vec{\rho}
\label{eq:cme.dir.5}
\end{eqnarray}
The Fourier transformation with the help of notations,
(\ref{eq:deh.2eh.15}) - (\ref{eq:deh.2eh.16}), easily reduces this
expression to Eq. (\ref{eq:deh.2eh.17}).

\subsection{The exchange part.}
\label{sec:cme.exc}

After the summation with respect to spin variables, Eq.
(\ref{eq:deh.2eh.11}) becomes
\begin{eqnarray}
    H_{\vec{K}_{1}s_{1},\vec{K}_{2}s_{2},
        \vec{K}_{3}s_{3},\vec{K}_{4}s_{4}}^{(x)} = & - &
    \delta_{s_{1e}s_{4e}}\delta_{s_{2e}s_{3e}}
    \delta_{s_{1h}s_{3h}}\delta_{s_{2h}s_{4h}}
    \int
    \psi_{\vec{K}_{1}}^{\ast}(\vec{r}_{e1}, \vec{r}_{h1})
    \psi_{\vec{K}_{2}}^{\ast}(\vec{r}_{e2}, \vec{r}_{h2})
\nonumber \\ && \times
    {\cal H}_{2eh}
    \psi_{\vec{K}_{3}}(\vec{r}_{e2}, \vec{r}_{h1})
    \psi_{\vec{K}_{4}}(\vec{r}_{e1}, \vec{r}_{h2})
    d\vec{r}_{e1} d\vec{r}_{h1} d\vec{r}_{e2} d\vec{r}_{h2}
\nonumber \\ & - &
    \delta_{s_{1e}s_{3e}}\delta_{s_{2e}s_{4e}}
    \delta_{s_{1h}s_{4h}}\delta_{s_{2h}s_{3h}}
    \int
    \psi_{\vec{K}_{1}}^{\ast}(\vec{r}_{e1}, \vec{r}_{h1})
    \psi_{\vec{K}_{2}}^{\ast}(\vec{r}_{e2}, \vec{r}_{h2})
\nonumber \\ && \times
    {\cal H}_{2eh}
    \psi_{\vec{K}_{4}}(\vec{r}_{e2}, \vec{r}_{h1})
    \psi_{\vec{K}_{3}}(\vec{r}_{e1}, \vec{r}_{h2})
    d\vec{r}_{e1} d\vec{r}_{h1} d\vec{r}_{e2} d\vec{r}_{h2} \ .
\label{eq:cme.exc.1}
\end{eqnarray}
The fact that these terms correspond to the exchange of two
electrons {\bf or} two holes, so the initial and final two
excitons are formed from different particles, makes the
calculation more complicated than the calculation of the direct
part. First, we cannot explicitly express products of the
electron and hole spin $\delta$-symbols in the exciton spin as we
did in the direct part, Eq. (\ref{eq:cme.dir.1}). Second, it is
harder to separate the single exciton energy $E_K$ from the
exciton-exciton interaction. Operating by Eq.
(\ref{eq:cme.dir.2}) on the wave functions
$\psi_{\vec{K}_{4}}(\vec{r}_{e1}, \vec{r}_{h2})
\psi_{\vec{K}_{3}}(\vec{r}_{e2}, \vec{r}_{h1})$ and
$\psi_{\vec{K}_{3}}(\vec{r}_{e1}, \vec{r}_{h2})
\psi_{\vec{K}_{4}}(\vec{r}_{e2}, \vec{r}_{h1})$ results in the
sum of single exciton energies of the initial state,
$E_{K_3}+E_{K_4}$. However, operating by the same kinetic terms
plus different terms of the Coulomb interaction, i.e.,
\begin{equation}
- {{\hbar^2 \nabla_{e1}^2}\over{2 m_e}}
    - {{\hbar^2 \nabla_{h1}^2}\over{2 m_\parallel}}+
    u_{eh}(|\vec{r}_{e1} - \vec{r}_{h1}|)
    - {{\hbar^2 \nabla_{e2}^2}\over{2 m_e}}
    - {{\hbar^2 \nabla_{h2}^2}\over{2 m_\parallel}}  +
    u_{eh}(|\vec{r}_{e2} - \vec{r}_{h2}|)
\end{equation}
on the wave function $\psi^{\ast}_{\vec{K}_{1}}(\vec{r}_{e1}, \vec{r}_{h1})
\psi^{\ast}_{\vec{K}_{2}}(\vec{r}_{e2}, \vec{r}_{h2})$ results in the sum of
single exciton energies of the final state, $E_{K_1}+E_{K_2}$. We write the
exchange part in a symmetric form so we add and subtract the kinetic terms
to the Hamiltonian and reduce Eq. (\ref{eq:cme.exc.1}) to
\begin{eqnarray}
H_{\vec{K}_{1}s_{1},\vec{K}_{2}s_{2},
        \vec{K}_{3}s_{3},\vec{K}_{4}s_{4}}^{(x)} = & - &
    \delta_{s_{1e}s_{4e}}\delta_{s_{2e}s_{3e}}
    \delta_{s_{1h}s_{3h}}\delta_{s_{2h}s_{4h}}
        \int
        \psi_{\vec{K}_{1}}^{\ast}(\vec{r}_{e1}, \vec{r}_{h1})
        \psi_{\vec{K}_{2}}^{\ast}(\vec{r}_{e2}, \vec{r}_{h2})
\nonumber \\ && \hspace{-3cm} \times
                \Bigg[
    E_{K_{1}} + E_{K_{2}} + E_{K_{3}} + E_{K_{4}} +
    {{\hbar^2 \nabla_{e1}^2}\over{2 m_e}} +
    {{\hbar^2 \nabla_{h1}^2}\over{2 m_\parallel}} +
    {{\hbar^2 \nabla_{e2}^2}\over{2 m_e}} +
    {{\hbar^2 \nabla_{h2}^2}\over{2 m_\parallel}}
\nonumber \\ && \hspace{-1cm} +
    u_{ee}(|\vec{r}_{e1} - \vec{r}_{e2}|) +
    u_{hh}(|\vec{r}_{h1} - \vec{r}_{h2}|)
                \Bigg]
\nonumber \\ && \hspace{-3cm} \times
        \psi_{\vec{K}_{3}}(\vec{r}_{e2}, \vec{r}_{h1})
        \psi_{\vec{K}_{4}}(\vec{r}_{e1}, \vec{r}_{h2})
        d\vec{r}_{e1} d\vec{r}_{h1} d\vec{r}_{e2} d\vec{r}_{h2}
\nonumber \\ & - &
    \delta_{s_{1e}s_{3e}}\delta_{s_{2e}s_{4e}}
    \delta_{s_{1h}s_{4h}}\delta_{s_{2h}s_{3h}}
        \int
        \psi_{\vec{K}_{1}}^{\ast}(\vec{r}_{e1}, \vec{r}_{h1})
        \psi_{\vec{K}_{2}}^{\ast}(\vec{r}_{e2}, \vec{r}_{h2})
\nonumber \\ && \hspace{-3cm} \times
                \Bigg[
    E_{K_{1}} + E_{K_{2}} + E_{K_{3}} + E_{K_{4}} +
    {{\hbar^2 \nabla_{e1}^2}\over{2 m_e}} +
    {{\hbar^2 \nabla_{h1}^2}\over{2 m_\parallel}} +
    {{\hbar^2 \nabla_{e2}^2}\over{2 m_e}} +
    {{\hbar^2 \nabla_{h2}^2}\over{2 m_\parallel}}
\nonumber \\ && \hspace{-1cm} +
    u_{ee}(|\vec{r}_{e1} - \vec{r}_{e2}|) +
    u_{hh}(|\vec{r}_{h1} - \vec{r}_{h2}|)
                \Bigg]
\nonumber \\ && \hspace{-3cm} \times
        \psi_{\vec{K}_{3}}(\vec{r}_{e1}, \vec{r}_{h2})
        \psi_{\vec{K}_{4}}(\vec{r}_{e2}, \vec{r}_{h1})
        d\vec{r}_{e1} d\vec{r}_{h1} d\vec{r}_{e2} d\vec{r}_{h2} \ .
\label{eq:cme.exc.2}
\end{eqnarray}
Convenient variables for the calculation of the integral are
coordinates $\vec{R}$, $\vec{r}_{ee}$ and $\vec{r}_{hh}$, Eq.
(\ref{eq:cme.oi.3}), and the center of mass of the system
$\vec{R}_c$, Eq. (\ref{eq:cme.dir.3a}). After integration with
respect to $\vec{R}_c$ we obtain
\begin{eqnarray}
H_{\vec{K}_{1}s_{1},\vec{K}_{2}s_{2},
        \vec{K}_{3}s_{3},\vec{K}_{4}s_{4}}^{(x)}  & = &
    {1 \over S} \ \delta_{\vec{K}_{1}+\vec{K}_{2},\vec{K}_{3}+\vec{K}_{4}}
        \left(
    E_{K_{1}} + E_{K_{2}} + E_{K_{3}} + E_{K_{4}} -
    {\hbar^{2} K^{2} \over 4M}
        \right)
\label{eq:cme.exc.3} \\ && \hspace{-2cm} \times
        \left(
    \delta_{s_{1e}s_{4e}}\delta_{s_{2e}s_{3e}}
    \delta_{s_{1h}s_{3h}}\delta_{s_{2h}s_{4h}}
    A_{\vec{K}_{1}\vec{K}_{2},\vec{K}_{3},\vec{K}_{4}} +
    \delta_{s_{1e}s_{3e}}\delta_{s_{2e}s_{4e}}
    \delta_{s_{1h}s_{4h}}\delta_{s_{2h}s_{3h}}
    A_{\vec{K}_{1}\vec{K}_{2},\vec{K}_{4},\vec{K}_{3}}
        \right)
\nonumber \\ & + &
    {1 \over S} \ \delta_{\vec{K}_{1}+\vec{K}_{2},\vec{K}_{3}+\vec{K}_{4}}
\nonumber \\ && \hspace{-2cm} \times
        \left(
    \delta_{s_{1e}s_{4e}}\delta_{s_{2e}s_{3e}}
    \delta_{s_{1h}s_{3h}}\delta_{s_{2h}s_{4h}}
    U_{\vec{K}_{1}\vec{K}_{2},\vec{K}_{3},\vec{K}_{4}}^{(x)} +
    \delta_{s_{1e}s_{3e}}\delta_{s_{2e}s_{4e}}
    \delta_{s_{1h}s_{4h}}\delta_{s_{2h}s_{3h}}
    U_{\vec{K}_{1}\vec{K}_{2},\vec{K}_{4},\vec{K}_{3}}^{(x)}
        \right) ,
\nonumber
\end{eqnarray}
where
\begin{eqnarray}
U_{\vec{K}_{1}\vec{K}_{2},\vec{K}_{3},\vec{K}_{4}}^{(x)} & = &
    - \int
    d\vec{R} d\vec{r}_{ee} d\vec{r}_{hh}
\nonumber \\ && \hspace{-1cm} \times
    \exp    \left[
    - i(\vec{K}_{1} - \vec{K}_{2}) \ {\vec{r}_{ee} + \vec{r}_{hh} \over 2}
            \right]
    \phi\left(\vec{R} + {\vec{r}_{ee} - \vec{r}_{hh} \over 2}\right)
    \phi\left(\vec{R} - {\vec{r}_{ee} - \vec{r}_{hh} \over 2}\right)
\nonumber \\ && \hspace{-1cm} \times
        \left[
    u_{ee}(r_{ee}) + u_{hh}(r_{hh}) +
    {\hbar^{2} \over 4\mu} \nabla^{2} +
    {\hbar^{2} \over m_{e}} \nabla_{ee}^{2} +
    {\hbar^{2} \over m_{h}} \nabla_{hh}^{2}
        \right]
\nonumber \\ && \hspace{-1cm} \times
    \exp    \left[
    - i(\vec{K}_{3} - \vec{K}_{4}) \ {\vec{r}_{ee} - \vec{r}_{hh} \over 2}
            \right]
    \phi\left(\vec{R} - {\vec{r}_{ee} + \vec{r}_{hh} \over 2}\right)
    \phi\left(\vec{R} + {\vec{r}_{ee} + \vec{r}_{hh} \over 2}\right) .
\label{eq:cme.exc.4}
\end{eqnarray}
From Eq. (\ref{eq:cme.exc.4}) it follows that
\begin{equation}
U_{\vec{K}_{1}\vec{K}_{2},\vec{K}_{3},\vec{K}_{4}}^{(x)} =
        U_{\vec{K}_{3},\vec{K}_{4},\vec{K}_{1}\vec{K}_{2}}^{(x)} \ .
\label{eq:cme.exc.5}
\end{equation}

The matrix element,
$U_{\vec{K}_{1}\vec{K}_{2},\vec{K}_{3},\vec{K}_{4}}^{(x)}$, is
simplified with the help of the low exciton energy assumption,
(\ref{eq:deh.2}). This is done in a similar way to the
simplification of
$A_{\vec{K}_{1}\vec{K}_{2},\vec{K}_{3},\vec{K}_{4}}$, in Appendix
\ref{sec:cme.oi}. The result does not depend on the excitons
momenta,
\begin{eqnarray}
U_{\vec{K}_{1}\vec{K}_{2},\vec{K}_{3},\vec{K}_{4}}^{(x)} & = &
    U^{(x)} \equiv - \int
    d\vec{R} d\vec{r}_{ee} d\vec{r}_{hh}
\nonumber \\ && \hspace{-1cm} \times
    \phi\left(\vec{R} + {\vec{r}_{ee} - \vec{r}_{hh} \over 2}\right)
    \phi\left(\vec{R} - {\vec{r}_{ee} - \vec{r}_{hh} \over 2}\right)
\nonumber \\ && \hspace{-1cm} \times
        \left[
    u_{ee}(r_{ee}) + u_{hh}(r_{hh}) +
    {\hbar^{2} \over 4\mu} \nabla^{2} +
    {\hbar^{2} \over m_{e}} \nabla_{ee}^{2} +
    {\hbar^{2} \over m_{h}} \nabla_{hh}^{2}
        \right]
\nonumber \\ && \hspace{-1cm} \times
    \phi\left(\vec{R} - {\vec{r}_{ee} + \vec{r}_{hh} \over 2}\right)
    \phi\left(\vec{R} + {\vec{r}_{ee} + \vec{r}_{hh} \over 2}\right) .
\label{eq:cme.exc.6}
\end{eqnarray}
This integral takes a more compact form if it is expressed in Fourier
transform of the wave function,
\begin{eqnarray}
U^{(x)} & = &
    {\hbar^{2} \over \mu}
    \int \phi_{q}^{4} \ q^{2} \ {d\vec{q} \over (2\pi)^{2}} -
    \int \phi_{q_{1}}^{2}\phi_{q_{2}}^{2} \
        \left[
    u_{ee}\left(|\vec{q}_{1} - \vec{q}_{2}|\right) +
    u_{hh}\left(|\vec{q}_{1} - \vec{q}_{2}|\right)
        \right]
    {d\vec{q}_{1}d\vec{q}_{2} \over (2\pi)^{4}} \ .
\label{eq:cme.exc.7}
\end{eqnarray}
This last integral can also be written in another form,
\begin{eqnarray}
U^{(x)} & = &
    {\hbar^{2} \over \mu}
    \int \phi_{q}^{4} \ q^{2} \ {d\vec{q} \over (2\pi)^{2}} -
    \int \left[u_{ee}(r) + u_{hh}(r)\right]
        \left[
    \int \phi_{q}^{2} e^{i\vec{q}\vec{r}} {d\vec{q} \over (2\pi)^{2}}
        \right]^{2} d\vec{r} \ .
\label{eq:cme.exc.8}
\end{eqnarray}

Integral (\ref{eq:cme.exc.6}) is of the order of $\hbar^{2}/\mu$. Since
$\epsilon_{b}\sim\hbar^{2}/\mu a^{2}$, the second term in
Eq. (\ref{eq:cme.exc.3}) is of the order of
$U^{(x)}/S\sim\epsilon_{b}(a^{2}/S)$. According to Eq. (\ref{eq:cme.oi.4}),
$A\sim a^{2}$, hence the first term in
Eq. (\ref{eq:cme.exc.3}) is of the same order as the second. The exciton kinetic
energy in the first term can be neglected because its contribution to
$H_{\vec{K}_{1}s_{1},\vec{K}_{2}s_{2},
        \vec{K}_{3}s_{3},\vec{K}_{4}s_{4}}^{(x)}$ is of the order
$\epsilon_{b}(a^{2}/S)(\hbar^{2}K^{2}/M)$, i.e., contains the product of two
small parameters. As a result Eq. (\ref{eq:cme.exc.3}) is reduced to
Eq. (\ref{eq:deh.2eh.18}).

\subsection{Contribution of excited states}
\label{sec:cme.exex}

For two excitons the definition, (\ref{eq:deh.rh.14}), can be
written as
\begin{eqnarray}
H_{\vec{K}_{1}s_{1},\vec{K}_{2}s_{2},
        \vec{K}_{3}s_{3},\vec{K}_{4}s_{4}}^{(exex)} & = &
\sum_{\vec{q}_{1},\vec{q}_{2}, \alpha_{1},\alpha_{2} \atop
    s^{\prime}_{1},s^{\prime}_{2}}
        \langle
    \vec{K}_{1}s_{1}\vec{K}_{2}s_{2}| {\cal H} + 2\epsilon_{b} |
    \vec{q}_{1}\alpha_{1}s^{\prime}_{1}\vec{q}_{2}\alpha_{2}s^{\prime}_{2}
        \rangle \
\label{eq:cme.exex.1}\\ && \hspace{-2cm} \times
        \left(
    E_{\alpha_{1}} + E_{\alpha_{2}} +
    {\hbar^{2}q_{1}^{2} \over 2M} + {\hbar^{2}q_{2}^{2} \over 2M} +
    2\epsilon_{b}
        \right)^{-1}
        \langle
    \vec{q}_{1}\alpha_{1}s^{\prime}_{1}\vec{q}_{2}\alpha_{2}s^{\prime}_{2}
    | {\cal H} + 2\epsilon_{b} |\vec{K}_{3}s_{3}\vec{K}_{4}s_{4}
        \rangle \ .
\nonumber
\end{eqnarray}
The matrix elements of the Hamiltonian between the ground state
and excited states can be separated into the direct and exchange
parts [compare Eqs. (\ref{eq:deh.2eh.9}) - (\ref{eq:deh.2eh.11})],
\begin{eqnarray}
&&        \langle
    \vec{K}_{1}s_{1}\vec{K}_{2}s_{2}
    | {\cal H} + 2\epsilon_{b} |
    \vec{q}_{1}\alpha_{1}s^{\prime}_{1}\vec{q}_{2}\alpha_{2}s^{\prime}_{2}
        \rangle = H^{(d)}_{\vec{q}_{1}\alpha_{1}s^{\prime}_{1}
    \vec{q}_{2}\alpha_{2}s^{\prime}_{2}}
    (\vec{K}_{1}s_{1},\vec{K}_{2}s_{2}) +
    H^{(d)}_{\vec{q}_{1}\alpha_{1}s^{\prime}_{1}
    \vec{q}_{2}\alpha_{2}s^{\prime}_{2}}
    (\vec{K}_{2}s_{2},\vec{K}_{1}s_{1})
\nonumber \\
&& \hspace{4cm}
    - H^{(x)}_{\vec{q}_{1}\alpha_{1}s^{\prime}_{1}
    \vec{q}_{2}\alpha_{2}s^{\prime}_{2}}
    (\vec{K}_{1}s_{1},\vec{K}_{2}s_{2}) -
    H^{(x)}_{\vec{q}_{1}\alpha_{1}s^{\prime}_{1}
    \vec{q}_{2}\alpha_{2}s^{\prime}_{2}}
    (\vec{K}_{2}s_{2},\vec{K}_{1}s_{1}) \ .
\label{eq:cme.exex.2}
\end{eqnarray}

In the calculation of these matrix elements it is convenient to
separate the sum of two single-exciton parts of the Hamiltonian,
(\ref{eq:cme.dir.2}), and to operate it on the left bracket
[similar to the calculation of
$H_{\vec{K}_{1}s_{1},\vec{K}_{2}s_{2},
        \vec{K}_{3}s_{3},\vec{K}_{4}s_{4}}^{(d)}$ and
$H_{\vec{K}_{1}s_{1},\vec{K}_{2}s_{2},
        \vec{K}_{3}s_{3},\vec{K}_{4}s_{4}}^{(x)}$]. Then, after the neglect
of the exciton kinetic energy,
\begin{mathletters}
\begin{eqnarray}
H^{(d)}_{\vec{q}_{1}\alpha_{1}s^{\prime}_{1}\vec{q}_{2}\alpha_{2}s^{\prime}_{2}}
    (\vec{K}_{1}s_{1},\vec{K}_{2}s_{2}) & = &
    \sum_{\sigma_{e1}\sigma_{e2}\sigma_{h1}\sigma_{h2}}
    \int d\vec{r}_{e1}d\vec{r}_{e2}d\vec{r}_{h1}d\vec{r}_{h2}
\nonumber \\ && \hspace{-2cm} \times
    \Psi_{\vec{K}_1, s_1}^{\ast}
        (\vec{r}_{e1},\sigma_{e1}; \vec{r}_{h1},\sigma_{h1})
    \Psi_{\vec{K}_2,s_2}^{\ast}
        (\vec{r}_{e2},\sigma_{e2}; \vec{r}_{h2},\sigma_{h2})
    U(\vec{r}_{e1},\vec{r}_{e2},\vec{r}_{h1},\vec{r}_{h2})
\nonumber \\ && \hspace{-2cm} \times
    \Psi_{\vec{q}_{1}\alpha_{1}s_{1}^{\prime}}
        (\vec{r}_{e1},\sigma_{e1}; \vec{r}_{h1},\sigma_{h1})
    \Psi_{\vec{q}_{2}\alpha_{2}s_{2}^{\prime}}
        (\vec{r}_{e2},\sigma_{e2}; \vec{r}_{h2},\sigma_{h2}) \ ,
\label{eq:cme.exex.3a} \\
H^{(x)}_{\vec{q}_{1}\alpha_{1}s^{\prime}_{1}
    \vec{q}_{2}\alpha_{2}s^{\prime}_{2}}
    (\vec{K}_{1}s_{1},\vec{K}_{2}s_{2}) & = &
    \sum_{\sigma_{e1}\sigma_{e2}\sigma_{h1}\sigma_{h2}}
    \int d\vec{r}_{e1}d\vec{r}_{e2}d\vec{r}_{h1}d\vec{r}_{h2}
\nonumber \\ && \hspace{-2cm} \times
    \Psi_{\vec{K}_1, s_1}^{\ast}
        (\vec{r}_{e1},\sigma_{e1}; \vec{r}_{h1},\sigma_{h1})
    \Psi_{\vec{K}_2,s_2}^{\ast}
        (\vec{r}_{e2},\sigma_{e2}; \vec{r}_{h2},\sigma_{h2})
    U(\vec{r}_{e1},\vec{r}_{e2},\vec{r}_{h1},\vec{r}_{h2})
\nonumber \\ && \hspace{-2cm} \times
    \Psi_{\vec{q}_{1}\alpha_{1}s_{1}^{\prime}}
        (\vec{r}_{e1},\sigma_{e1}; \vec{r}_{h2},\sigma_{h2})
    \Psi_{\vec{q}_{2}\alpha_{2}s_{2}^{\prime}}
        (\vec{r}_{e2},\sigma_{e2}; \vec{r}_{h1},\sigma_{h1}) \ ,
\label{eq:cme.exex.3b}
\end{eqnarray}
\label{eq:cme.exex.3}
\end{mathletters}
where
\begin{eqnarray}
U(\vec{r}_{e1},\vec{r}_{e2},\vec{r}_{h1},\vec{r}_{h2}) & = &
    u_{ee}(|\vec{r}_{e1} - \vec{r}_{e2}|) +
    u_{hh}(|\vec{r}_{h1} - \vec{r}_{h2}|)
\nonumber \\ & + &
    u_{eh}(|\vec{r}_{e1} - \vec{r}_{h2}|) +
    u_{eh}(|\vec{r}_{e2} - \vec{r}_{h1}|) \ .
\label{eq:cme.exex.4}
\end{eqnarray}

After the substitution of the exciton wave function, Eqs.
(\ref{eq:deh.rh.4}) - (\ref{eq:deh.rh.6}), in Eq.
(\ref{eq:cme.exex.3}), the spin factors are easily separated,
\begin{mathletters}
\begin{eqnarray}
H^{(d)}_{\vec{q}_{1}\alpha_{1}s^{\prime}_{1}
    \vec{q}_{2}\alpha_{2}s^{\prime}_{2}}
    (\vec{K}_{1}s_{1},\vec{K}_{2}s_{2}) & = &
    \delta_{s_{1},s_{e1}^{\prime}+s_{h1}^{\prime}}
    \delta_{s_{2},s_{e2}^{\prime}+s_{h2}^{\prime}}
    H^{(d)}_{\vec{q}_{1}\alpha_{1},\vec{q}_{2}\alpha_{2}}
    (\vec{K}_{1},\vec{K}_{2})
\label{eq:cme.exex.5a} \\
H^{(x)}_{\vec{q}_{1}\alpha_{1}s^{\prime}_{1}
    \vec{q}_{2}\alpha_{2}s^{\prime}_{2}}
    (\vec{K}_{1}s_{1},\vec{K}_{2}s_{2}) & = &
    \delta_{s_{1},s_{e1}^{\prime}+s_{h2}^{\prime}}
    \delta_{s_{2},s_{e2}^{\prime}+s_{h1}^{\prime}}
    H^{(x)}_{\vec{q}_{1}\alpha_{1},\vec{q}_{2}\alpha_{2}}
    (\vec{K}_{1},\vec{K}_{2}) \ ,
\label{eq:cme.exex.5b}
\end{eqnarray}
\label{eq:cme.exex.5}
\end{mathletters}
and
\begin{mathletters}
\begin{eqnarray}
&& H^{(d)}_{\vec{q}_{1}\alpha_{1},\vec{q}_{2}\alpha_{2}}
    (\vec{K}_{1},\vec{K}_{2}) =
    {1 \over S^{2}}
    \int d\vec{r}_{e1}d\vec{r}_{e2}d\vec{r}_{h1}d\vec{r}_{h2}
    U(\vec{r}_{e1},\vec{r}_{e2},\vec{r}_{h1},\vec{r}_{h2})
\label{eq:cme.exex.6a}  \\
&&  \hspace{2cm} \times  \exp    \left[
    - i\vec{K}_{1} {m_{e}\vec{r}_{e1} + m_{\parallel}\vec{r}_{h1} \over M}
    - i\vec{K}_{2} {m_{e}\vec{r}_{e2} + m_{\parallel}\vec{r}_{h2} \over M}
            \right]
    \phi(|\vec{r}_{e1} - \vec{r}_{h1}|) \phi(|\vec{r}_{e2} - \vec{r}_{h2}|)
\nonumber \\ && \hspace{2cm} \times
    \exp    \left[
    i\vec{q}_{1} {m_{e}\vec{r}_{e1} + m_{\parallel}\vec{r}_{h1} \over M} +
    i\vec{q}_{2} {m_{e}\vec{r}_{e2} + m_{\parallel}\vec{r}_{h2} \over M}
            \right]
    \phi_{\alpha_{1}}(\vec{r}_{e1} - \vec{r}_{h1})
    \phi_{\alpha_{2}}(\vec{r}_{e2} - \vec{r}_{h2}) \ ,
\nonumber \\
&& H^{(x)}_{\vec{q}_{1}\alpha_{1},\vec{q}_{2}\alpha_{2}}
    (\vec{K}_{1},\vec{K}_{2}) =
    {1 \over S^{2}}
    \int d\vec{r}_{e1}d\vec{r}_{e2}d\vec{r}_{h1}d\vec{r}_{h2} \
    U(\vec{r}_{e1},\vec{r}_{e2},\vec{r}_{h1},\vec{r}_{h2})
\label{eq:cme.exex.6b} \\
&&   \hspace{2cm} \times \exp    \left[
    - i\vec{K}_{1} {m_{e}\vec{r}_{e1} + m_{\parallel}\vec{r}_{h1} \over M}
    - i\vec{K}_{2} {m_{e}\vec{r}_{e2} + m_{\parallel}\vec{r}_{h2} \over M}
            \right]
    \phi(|\vec{r}_{e1} - \vec{r}_{h1}|) \phi(|\vec{r}_{e2} - \vec{r}_{h2}|)
\nonumber \\
&& \hspace{2cm} \times
    \exp    \left[
    i\vec{q}_{1} {m_{e}\vec{r}_{e1} + m_{\parallel}\vec{r}_{h2} \over M} +
    i\vec{q}_{2} {m_{e}\vec{r}_{e2} + m_{\parallel}\vec{r}_{h1} \over M}
            \right]
    \phi_{\alpha_{1}}(\vec{r}_{e1} - \vec{r}_{h2})
    \phi_{\alpha_{2}}(\vec{r}_{e2} - \vec{r}_{h1}) \ .
\nonumber
\end{eqnarray}
\label{eq:cme.exex.6}
\end{mathletters}
The integrals are simplified with the help of new integration
variables. For the direct part they are given by Eq.
(\ref{eq:cme.dir.3}) and for the exchange part they are the
two-exciton center of mass, $\vec{R}_{c}$, Eq.
(\ref{eq:cme.dir.3a}), and the relative coordinates, Eq.
(\ref{eq:cme.oi.3}). After the integration with respect to
$\vec{R}_{c}$
\begin{mathletters}
\begin{eqnarray}
H^{(d)}_{\vec{q}_{1}\alpha_{1},\vec{q}_{2}\alpha_{2}}
    (\vec{K}_{1},\vec{K}_{2}) & = &
    {1 \over S} \
    \delta_{\vec{K}_{1} + \vec{K}_{2}, \vec{q}_{1} + \vec{q}_{2}} \
    D_{\alpha_{1}\alpha_{2}}^{(d)}
    (\vec{K}_{1},\vec{K}_{2},\vec{q}_{1},\vec{q}_{2}) \ ,
\label{eq:cme.exex.7a} \\
H^{(x)}_{\vec{q}_{1}\alpha_{1},\vec{q}_{2}\alpha_{2}}
    (\vec{K}_{1},\vec{K}_{2}) & = &
    {1 \over S} \
    \delta_{\vec{K}_{1} + \vec{K}_{2}, \vec{q}_{1} + \vec{q}_{2}} \
    D_{\alpha_{1}\alpha_{2}}^{(x)}
    (\vec{K}_{1},\vec{K}_{2},\vec{q}_{1},\vec{q}_{2}) \ .
\label{eq:cme.exex.7b}
\end{eqnarray}
\label{eq:cme.exex.7}
\end{mathletters}

In the remaining integrals, $D_{\alpha_{1}\alpha_{2}}^{(d)}$ and
$D_{\alpha_{1}\alpha_{2}}^{(x)}$, it is possible to neglect
$K_{1}$ and $K_{2}$ which due to $\delta$-symbols in Eq.
(\ref{eq:cme.exex.7}) means also that $\vec{q}_{2}=-\vec{q}_{1}$.
This allows us to simplify the notations,
$D_{\alpha_{1}\alpha_{2}}^{(d)}
(\vec{K}_{1},\vec{K}_{2},\vec{q}_{1},\vec{q}_{2})\equiv
D_{\alpha_{1}\alpha_{2}}^{(d)}(q_{1})$,
$D_{\alpha_{1}\alpha_{2}}^{(x)}
(\vec{K}_{1},\vec{K}_{2},\vec{q}_{1},\vec{q}_{2})\equiv
D_{\alpha_{1}\alpha_{2}}^{(x)}(q_{1})$. For further
simplification it is convenient to substitute the Fourier
transform of the exciton functions,
\begin{equation}
\phi_{\alpha,\vec{q}} = \int e^{- i\vec{q}\vec{r}} \phi_{\alpha}(\vec{r})
    d\vec{r} \ .
\label{eq:cme.exex.8}
\end{equation}
The calculations are a bit cumbersome but straightforward and
they lead to Eqs. (\ref{eq:deh.2eh.21}) - (\ref{eq:deh.2eh.23}).

\section{Spin sums in the exchange term}
\label{xsr}

In this appendix we reduce the spin sum in the exchange term of
the many exciton Hamiltonian, Eq. (\ref{eq:deh.heg.6}), to a
simpler form. According to the definition, $s_{j}=s_{ej}+s_{hj}$,
and there is one to one correspondence between the electron and
hole spins on one side and the exciton spin on the other.
Therefore
\begin{eqnarray}
&& \sum_{s_{1},s_{2},s_{3},s_{4}}
    \left(\delta_{s_{e1},s_{e4}} \delta_{s_{e2},s_{e3}}
    \delta_{s_{h1},s_{h3}} \delta_{s_{h2},s_{h4}} +
    \delta_{s_{e1},s_{e3}} \delta_{s_{e2},s_{e4}}
    \delta_{s_{h1},s_{h4}} \delta_{s_{h2},s_{h3}}\right)
\nonumber \\
&&  \times
    c^\dag_{\vec{K}_2,s_2}  c^\dag_{\vec{K}_1,s_1}
    c_{\vec{K}_1-\vec{q},s_3} c_{\vec{K}_2+\vec{q},s_4} .
\nonumber \\
&&  = \sum_{s_{e1},s_{e2},s_{e3},s_{e4},s_{h1},s_{h2},s_{h3},s_{h4}}
    \left(\delta_{s_{e1},s_{e4}} \delta_{s_{e2},s_{e3}}
    \delta_{s_{h1},s_{h3}} \delta_{s_{h2},s_{h4}} +
    \delta_{s_{e1},s_{e3}} \delta_{s_{e2},s_{e4}}
    \delta_{s_{h1},s_{h4}} \delta_{s_{h2},s_{h3}}\right)
\nonumber \\
&&  \times
    c^\dag_{\vec{K}_2,s_{e2}+s_{h2}}  c^\dag_{\vec{K}_1,s_{e1}+s_{h1}}
    c_{\vec{K}_1-\vec{q},s_{e3}+s_{h3}} c_{\vec{K}_2+\vec{q},s_{e4}+s_{h4}}=
\label{eq:xsr.1} \\
&&  =\sum_{s_{e1},s_{h1},s_{e2},s_{h2}}
    c^\dag_{\vec{K}_2,s_{e2}+s_{h2}}  c^\dag_{\vec{K}_1,s_{e1}+s_{h1}}
    \left(c_{\vec{K}_1-\vec{q},s_{e2}+s_{h1}} c_{\vec{K}_2+\vec{q},s_{e1}+s_{h2}}+
    c_{\vec{K}_1-\vec{q},s_{e1}+s_{h2}} c_{\vec{K}_2+\vec{q},s_{e2}+s_{h1}}\right)
\nonumber \\
&&  =\Sigma_1+\Sigma_2+\Sigma_3+\Sigma_4 \ ,
\nonumber
\end{eqnarray}
where
\begin{mathletters}
\begin{eqnarray}
&&  \Sigma_1=2 \sum_{s_{e},s_{h}}
    c^\dag_{\vec{K}_2,s_{e}+s_{h}}  c^\dag_{\vec{K}_1,s_{e}+s_{h}}
    c_{\vec{K}_1-\vec{q},s_{e}+s_{h}} c_{\vec{K}_2+\vec{q},s_{e}+s_{h}} \ ,
\label{eq:xsr.2a} \\
&&  \Sigma_2 =
    \sum_{s_{e},s_{h}} c^\dag_{\vec{K}_2,s_{e}+s_{h}}  c^\dag_{\vec{K}_1,-s_{e}+s_{h}}
    \left(c_{\vec{K}_1-\vec{q},-s_{e}+s_{h}} c_{\vec{K}_2+\vec{q},s_{e}+s_{h}}+
    c_{\vec{K}_1-\vec{q},s_{e}+s_{h}} c_{\vec{K}_2+\vec{q},-s_{e}+s_{h}}\right) \ ,
\label{eq:xsr.2b} \\
&&  \Sigma_3 =\sum_{s_{e},s_{h}} c^\dag_{\vec{K}_2,s_{e}+s_{h}}
c^\dag_{\vec{K}_1,s_{e}-s_{h}}
    \left(c_{\vec{K}_1-\vec{q},s_{e}+s_{h}} c_{\vec{K}_2+\vec{q},s_{e}-s_{h}}+
    c_{\vec{K}_1-\vec{q},s_{e}-s_{h}} c_{\vec{K}_2+\vec{q},s_{e}+s_{h}}\right) \ ,
\label{eq:xsr.2c} \\
&&  \Sigma_4=\sum_{s_{e},s_{h}} c^\dag_{\vec{K}_2,s_{e}+s_{h}}
c^\dag_{\vec{K}_1,-s_{e}-s_{h}}
    \left(c_{\vec{K}_1-\vec{q},-s_{e}+s_{h}} c_{\vec{K}_2+\vec{q},s_{e}-s_{h}}+
    c_{\vec{K}_1-\vec{q},s_{e}-s_{h}} c_{\vec{K}_2+\vec{q},-s_{e}+s_{h}}\right) \ .
\label{eq:xsr.2d}
\end{eqnarray}
\label{eq:xsr.2}
\end{mathletters}
Here we use the fact that the both the spin of the hole and the
spin of the electron have only two values, $s_{e}=\pm 1/2$,
$s_{h}=\pm 3/2$. Hence, the only possible values of $s_{h2}$ are
$\pm s_{h1}$ and of $s_{e2}$ are $\pm s_{e1}$. $\Sigma_i$
describe all the possible combinations that appear in Eq.
(\ref{eq:xsr.1}). The next step is to replace the quantum numbers
of the electron and hole spin with the quantum numbers of the
excitons spin. Here we use the one to one correspondence between
the electron and hole spins and the exciton spin
\begin{mathletters}
\begin{eqnarray}
&& \Sigma_1=2 \sum_{s} c^\dag_{\vec{K}_2,s} c^\dag_{\vec{K}_1,s}
    c_{\vec{K}_1-\vec{q},s} c_{\vec{K}_2+\vec{q},s} \ ,
\label{eq:xsr.3a} \\
&&  \Sigma_2+\Sigma_3 =
    \sum_{s_1,s_2} c^\dag_{\vec{K}_2,s_2}c^\dag_{\vec{K}_1,s_1}
        \left(
    c_{\vec{K}_1-\vec{q},s_1} c_{\vec{K}_2+\vec{q},s_2} +
    c_{\vec{K}_1-\vec{q},s_2} c_{\vec{K}_2+\vec{q},s_1}
        \right)
\nonumber \\ &&  -
    2 \sum_{s} c^\dag_{\vec{K}_2,s}  c^\dag_{\vec{K}_1,s}
    c_{\vec{K}_1-\vec{q},s} c_{\vec{K}_2+\vec{q},s} -
    \sum_{s} c^\dag_{\vec{K}_2,s}c^\dag_{\vec{K}_1,-s}
        \left(
    c_{\vec{K}_1-\vec{q},-s} c_{\vec{K}_2+\vec{q},s}+
    c_{\vec{K}_1-\vec{q},s} c_{\vec{K}_2+\vec{q},-s}
        \right)
    \ ,
\label{eq:xsr.3b}\\
&&  \Sigma_4=\sum_{s_1,s_2}
     c^\dag_{\vec{K}_2,s_1}  c^\dag_{\vec{K}_1,-s_1}
    c_{\vec{K}_1-\vec{q},-s_2} c_{\vec{K}_2+\vec{q},s_2}
\nonumber \\
&&  -\sum_{s} c^\dag_{\vec{K}_2,s}  c^\dag_{\vec{K}_1,-s}
    \left(c_{\vec{K}_1-\vec{q},-s} c_{\vec{K}_2+\vec{q},s}+
    c_{\vec{K}_1-\vec{q},-s} c_{\vec{K}_2+\vec{q},s}\right) \ .
\label{eq:xsr.3c}
\end{eqnarray}
\label{eq:xsr.3}
\end{mathletters}
The exchange term in the many exciton Hamiltonian, Eq. (\ref{eq:deh.heg.6}),
has now the following form
\begin{equation}
H^{(x)}={V_{x} \over 4 S}
    \sum_{\vec{K}_1,\vec{K}_2,\vec{q}}\left(\Sigma_1+\Sigma_2+
    \Sigma_3+\Sigma_4 \right) ,
\label{eq:xsr.4}
\end{equation}
Now we replace the summation over $\vec{q}$ with the summation over
$Q=\vec{K}_1-\vec{K}_2-\vec{q}$ in the second term in both parentheses in
$\Sigma_2+\Sigma_3$ and see that it becomes equivalent to the first term in
the same parentheses. The same can be done with the two terms inside the
parentheses in $\Sigma_4$. So the exchange term is
\begin{eqnarray}
&&  H^{(x)}={V_{x} \over 4 S}
    \sum_{\vec{K}_1,\vec{K}_2,\vec{q}}\left[2 \sum_{s_1,s_2}
     c^\dag_{\vec{K}_2,s_2}  c^\dag_{\vec{K}_1,s_1}
    c_{\vec{K}_1-\vec{q},s_1} c_{\vec{K}_2+\vec{q},s_2} \right.
\label{eq:xsr.5} \\
&&  \left. -4\sum_{s}  c^\dag_{\vec{K}_2,-s}  c^\dag_{\vec{K}_1,s}
    c_{\vec{K}_1-\vec{q},s} c_{\vec{K}_2+\vec{q},-s}
    +\sum_{s_1,s_2}
     c^\dag_{\vec{K}_2,-s_1}  c^\dag_{\vec{K}_1,s_1}
    c_{\vec{K}_1-\vec{q},s_2} c_{\vec{K}_2+\vec{q},-s_2} \right] \ .
\nonumber
\end{eqnarray}

\section{Fourier components of the interaction potential and the wave
function in quantum wells}
\label{sec:fc}

In this appendix we present some formulae necessary for the calculation of
Hamiltonian matrix elements.

The expressions for $u_{ij}(r)$ that are obtained with the help of
Eqs. (\ref{eq:Hehg.6}) and (\ref{eq:mfa.5}) are quite complicated. But their
Fourier components have a relatively simple form,
\begin{mathletters}
\begin{eqnarray}
u_{eh}(q) & = & -{2 \pi e^2 \over q \kappa}
    {(8 \pi^2)^2 \over L_h L_e}
    {{\sinh{(q L_e/2)} \sinh{(q L_h/2)}} \over
    {q^2 (4 \pi^2 + L_e^2 q^2)(4 \pi^2 + L_h^2 q^2)}} \
    e^{- q (w+L_h/2+L_e/2)} \ ,
\label{eq:eq:fc.1a} \\
u_{hh}(q) & = & {2 \pi e^2 \over q \kappa} \left[
    \displaystyle {2 \over q L_h} +
    {q L_h \over{4 \pi^2 + q^2 L_h^2}}-
    {{32 \pi^4 (1-\exp{(-q L_h)})}
    \over {q^2 L_h^2 ({4 \pi^2 + q^2 L_h^2})^2}}
    \right] \ ,
\label{eq:eq:fc.1b}\\
u_{ee}(q) & = & {2 \pi e^2 \over q \kappa} \left[
    \displaystyle {2 \over q L_e} +
    {q L_e \over{4 \pi^2 + q^2 L_e^2}}-
    {{32 \pi^4 (1-\exp{(-q L_e)})}
    \over {q^2 L_e^2 ({4 \pi^2 + q^2 L_e^2})^2}}
    \right] \ .
\label{eq:eq:fc.1c}
\end{eqnarray}
\label{eq:fc.1}
\end{mathletters}
Each of these functions is singular at $q=0$. But the direct interaction
potential which contains the sum of them is finite,
\begin{eqnarray}
U^{(d)}(0) & = &
    {2\pi e^{2} \over \kappa}
        \left[
    2w + \left({2 \over 3} + {5 \over 4\pi^{2}}\right) (L_{e} + L_{h})
        \right] .
\label{eq:fc.2}
\end{eqnarray}
This expression leads to Eq. (\ref{eq:mfa.7}).

The Fourier transform of single-exciton wave function,
(\ref{eq:mfa.6}), necessary for the calculation of $U^{(d)}(q)$
and $V_{x}$, is
\begin{eqnarray}
\phi_{q} = \sqrt{2\pi \over b(b + r_{0})} e^{r_{0}/2b} {1\over 2b}
    \left({1\over 4b^{2}} + q^{2}\right)^{-3/2}
    \left(1 + r_{0} \sqrt{{1\over 4b^{2}} + q^{2}}\right)
    \exp\left[- r_{0} \sqrt{{1\over 4b^{2}} + q^{2}}\right] \ .
\label{eq:fc.3}
\end{eqnarray}

\end{document}